\documentclass[zpreprint,zbstdefault]{zeus_paper}
\usepackage[english]{babel}

\newcommand{\ZcoosysB}{%
The ZEUS coordinate system is a right-handed Cartesian system, with the $Z$
axis pointing in the proton beam direction, referred to as the ``forward
direction'', and the $X$ axis pointing left towards the centre of HERA.
The coordinate origin is at the nominal interaction point.\xspace}

\newcommand{\ZcoosysfnB}{\footnote{\ZcoosysB}}

\newcommand{\Zctddesc}[1]{%
Charged particles were tracked in the central tracking detector (CTD)~\citeCTD,
which operated in a magnetic field of $1.43\Tesla$ provided by a thin 
superconducting coil. The CTD consisted of 72~cylindrical drift chamber 
layers, organised in 9~superlayers covering the polar-angle#1 region 
\mbox{$15^\circ<\theta<164^\circ$}. The transverse-momentum resolution for
full-length tracks was $\sigma(p_T)/p_T=0.0058p_T\oplus0.0065\oplus0.0014/p_T$,
with $p_T$ in $\Gev$.}
\newcommand{\Zcaldesc}{%
The high-resolution uranium--scintillator calorimeter (CAL)~\citeCAL consisted 
of three parts: the forward (FCAL), the barrel (BCAL) and the rear (RCAL)
calorimeters. Each part was subdivided transversely into towers and
longitudinally into one electromagnetic section (EMC) and either one (in RCAL)
or two (in BCAL and FCAL) hadronic sections (HAC). The smallest subdivision of
the calorimeter was called a cell.  The CAL energy resolutions, as measured under
test-beam conditions, were $\sigma(E)/E=0.18/\sqrt{E}$ for electrons and
$\sigma(E)/E=0.35/\sqrt{E}$ for hadrons, with  $E$ in $\Gev$.}

\chardef\usc=95
\chardef\til=126
\catcode`\@=11 
\DeclareRobustCommand\xdotspace{\futurelet\@let@token\@xdotspace}
\def\@xdotspace{%
  \ifx\@let@token.\else
  \ifx\@let@token\bgroup.\else
  \ifx\@let@token\egroup.\else
  \ifx\@let@token\/.\else
  \ifx\@let@token\ .\else
  \ifx\@let@token~.\else
  \ifx\@let@token!.\else
  \ifx\@let@token,.\else
  \ifx\@let@token:.\else
  \ifx\@let@token;.\else
  \ifx\@let@token?.\else
  \ifx\@let@token/.\else
  \ifx\@let@token'.\else
  \ifx\@let@token).\else
  \ifx\@let@token-.\else
  \ifx\@let@token\@xobeysp.\else
  \ifx\@let@token\space.\else
  \ifx\@let@token\@sptoken.\else
   .\space
   \fi\fi\fi\fi\fi\fi\fi\fi\fi\fi\fi\fi\fi\fi\fi\fi\fi\fi}
\catcode`\@=12 

\newcommand{\stru}[2]{%
   \relax\ifmmode\hbox{\vrule height#1 depth#2 width0pt}%
   \else\vrule height#1 depth#2 width0pt\fi}

\newcommand{\Ronum}[1]{\uppercase\expandafter{\romannumeral#1}}
\newcommand{\ronum}[1]{\expandafter{\romannumeral#1}}
\DeclareRobustCommand{\LaTeXZ}{%
  \LaTeX\kern-.05em4\kern-.1em
  {\raisebox{-0.2ex}{$\scriptstyle\text{ZEUS}$}}\xspace}

\DeclareMathAlphabet{\mathbf}{OT1}{cmr}{bx}{sl}
\newcommand{\eVdist}{\kern-0.06667em}

\newcommand{\Gev}{{\text{Ge}\eVdist\text{V\/}}}

\newcommand{\gev}{{\,\text{Ge}\eVdist\text{V\/}}}

\newcommand{\cm}{\,\text{cm}}

\newcommand{\Tesla}{\,\text{T}}

\newcommand{\slashfrac}[2]{%
  \raisebox{0.5ex}{\ensuremath #1}\kern-0.12em/\kern-0.08em
  \raisebox{-.8ex}{\ensuremath #2}}

\newcommand{\sqr}[3]{%
    {\vcenter{\hrule height.#3ex\hbox{\vrule width.#2ex height#1ex
     \kern#1ex\vrule width.#3ex}\hrule height.#2ex}}}

\catcode`\@=11 
\newcommand{\parenbar}{\mathpalette\p@renb@r}
\def\p@renb@r#1#2{\vbox{%
  \ifx#1\scriptscriptstyle \dimen@.7em\dimen@ii.2em\else
  \ifx#1\scriptstyle \dimen@.8em\dimen@ii.25em\else
  \dimen@1em\dimen@ii.4em\fi\fi \offinterlineskip
  \ialign{\hfill##\hfill\cr
    \vbox{\hrule width\dimen@ii}\cr
    \noalign{\vskip-.3ex}%
    \hbox to\dimen@{$\mathchar300\hfil\mathchar301$}\cr
    \noalign{\vskip-.3ex}%
    $#1#2$\cr}}}
\catcode`\@=12 

\newcommand{\IP}{{\rm I$\kern-0.01667em$P}\xspace}

\mathchardef\qsm=63
\mathchardef\pls=43
\mathchardef\mns=512
\mathchardef\plm=518
\mathchardef\eql=61
\mathchardef\smallleft=300
\mathchardef\smallright=301
\mathchardef\les=316
\mathchardef\gre=318
\mathchardef\leq=532
\mathchardef\grq=533

\catcode`\@=11 
\newcounter{pict@width}
\newcounter{pict@height}
\newlength{\pict@scale}
\newcommand{\psfigadd}[4]{%
\setcounter{pict@width}{1*\ratio{#2+\pict@scale/2}{\pict@scale}}
\setcounter{pict@height}{1*\ratio{#3+\pict@scale/2}{\pict@scale}}
\setlength{\unitlength}{\pict@scale}
\hbox to #2{\hspace{-\fill}\begin{picture}(\thepict@width,\thepict@height)
\put(0,0){\psfig{figure=#1,width=#2,height=#3,clip=}}
\SetScale{0.283466457}
\SetWidth{1.763889}
{#4}
\end{picture}}
}
\newcounter{pict@widthfst}
\newcounter{pict@widthscd}
\newcounter{pict@widthtot}
\newcommand{\psfigaddtwo}[7]{%
\setcounter{pict@widthfst}{1*\ratio{#2+\pict@scale/2}{\pict@scale}}
\setcounter{pict@widthscd}{1*\ratio{#2+#4+\pict@scale/2}{\pict@scale}}
\setcounter{pict@widthtot}{1*\ratio{#2+#4+#6+\pict@scale/2}{\pict@scale}}
\setcounter{pict@height}{1*\ratio{#3+\pict@scale/2}{\pict@scale}}
\setlength{\unitlength}{\pict@scale}
\hbox{\hspace{-\fill}\begin{picture}(\thepict@widthtot,\thepict@height)
\put(0,0){\psfig{figure=#1,width=#2,height=#3,clip=}}
\put(\thepict@widthscd,0){\psfig{figure=#5,width=#6,height=#3,clip=}}
\SetScale{0.283466457}
\SetWidth{1.763889}
{#7}
\end{picture}}
}
\newcommand{\psfigror}[4]{%
\setcounter{pict@width}{1*\ratio{#2+\pict@scale/2}{\pict@scale}}
\setcounter{pict@height}{1*\ratio{#3+\pict@scale/2}{\pict@scale}}
\setlength{\unitlength}{\pict@scale}
\hbox{\begin{picture}(\thepict@width,\thepict@height)
\put(0,\thepict@height){\psfig{figure=#1,width=#3,height=#2,clip=,angle=270}}
\SetScale{0.283466457}
\SetWidth{1.763889}
{#4}
\end{picture}}
}
\newcommand{\psfigrol}[4]{%
\setcounter{pict@width}{1*\ratio{#2+\pict@scale/2}{\pict@scale}}
\setcounter{pict@height}{1*\ratio{#3+\pict@scale/2}{\pict@scale}}
\setlength{\unitlength}{\pict@scale}
\hbox{\begin{picture}(\thepict@width,\thepict@height)
\put(0,0){\psfig{figure=#1,width=#3,height=#2,clip=,angle=90}}
\SetScale{0.283466457}
\SetWidth{1.763889}
{#4}
\end{picture}}
}
\catcode`\@=12 
\newlength\listtextwidth

\catcode`\@=11 
\newlength{\@tabfninsert}
\newlength{\@tabfnwidth}
\newcommand{\tabfootnote}[2]{%
  \setlength{\@tabfninsert}{0.8em}
  \setlength{\@tabfnwidth}{\textwidth}
  \addtolength{\@tabfnwidth}{-\@tabfninsert}
  \addtolength{\@tabfnwidth}{-0.4em}
  \noindent\makebox[\@tabfninsert][r]{\footnotesize$^{#1}$\hfil}\hfill%
  \parbox[t]{\@tabfnwidth}{\footnotesize #2\hfill}}
\catcode`\@=12 

\def\q2{Q^2}
\def\epem{{e^+e^-}}
\def\meff{M_{\rm eff}}
\def\nch{{\langle n_{\rm ch} \rangle}}
\def\twonch{{2\cdot \langle n_{\rm ch} \rangle}}
\def\ecrbreit{{E^{\rm cr}_{\rm B}}}
\def\twoecrbreit{{2\cdot E^{\rm cr}_{\rm B}}}

\def\citeCTD{{\cite{%
nim:a279:290,*npps:b32:181,*nim:a338:254%
}}\xspace}
\def\citeCAL{{\cite{%
nim:a309:77,*nim:a309:101,*nim:a321:356,*nim:a336:23%
}}\xspace}

\begin{document}
\prepnum{DESY--08--036}

\title{
Energy dependence of the 
charged multiplicity 
 in deep inelastic scattering at HERA
}                                                       
                    
\author{ZEUS Collaboration}
\draftversion{After reading}
\date{\today}

\abstract{
The charged multiplicity distributions and the mean charged multiplicity
have been investigated in inclusive
neutral current deep inelastic $ep$ scattering with the ZEUS detector
at HERA, using an integrated luminosity of 38.6~pb$^{-1}$. 
The measurements were performed in 
the current region
 of the Breit frame, as well as in the current
fragmentation region of the hadronic centre-of-mass frame.
The KNO-scaling properties of the data were investigated and the
energy dependence was studied 
 using different energy scales.
The data are compared 
to results obtained in $\epem$ collisions and to previous DIS measurements
as well as 
to leading-logarithm parton-shower Monte Carlo predictions.
}

\makezeustitle

%
%
%
%
\def\3{\ss}                                                                                        
\pagenumbering{Roman}                                                                              
                                                   %
\begin{center}                                                                                     
{                      \Large  The ZEUS Collaboration              }                               
\end{center}                                                                                       
  S.~Chekanov,                                                                                     
  M.~Derrick,                                                                                      
  S.~Magill,                                                                                       
  B.~Musgrave,                                                                                     
  D.~Nicholass$^{   1}$,                                                                           
  \mbox{J.~Repond},                                                                                
  R.~Yoshida\\                                                                                     
 {\it Argonne National Laboratory, Argonne, Illinois 60439-4815, USA}~$^{n}$                       
\par \filbreak                                                                                     
  M.C.K.~Mattingly \\                                                                              
 {\it Andrews University, Berrien Springs, Michigan 49104-0380, USA}                               
\par \filbreak                                                                                     
  M.~Jechow,                                                                                       
  N.~Pavel~$^{\dagger}$\\                                                                          
  {\it Institut f\"ur Physik der Humboldt-Universit\"at zu Berlin,                                 
           Berlin, Germany}~$^{b}$
% ...
\newpage                                                                                           
\enlargethispage{5cm}                                                                              
$^{\    1}$ also affiliated with University College London, UK \\                                  
% ...
$^{\    5}$ This work was supported in part by the Marie Curie Actions Transfer of Knowledge       
project COCOS (contract MTKD-CT-2004-517186)\\                                                     
% ...
$^{\dagger}$ deceased \\                                                                           
%
\newpage   
                                                           %
                                                           %
\begin{tabular}[h]{rp{14cm}}                                                                       
$^{a}$ &  supported by the Natural Sciences and Engineering Research Council of Canada (NSERC) \\  
$^{b}$ &  supported by the German Federal Ministry for Education and Research (BMBF), under        
          contract numbers 05 HZ6PDA, 05 HZ6GUA, 05 HZ6VFA and 05 HZ4KHA\\                         
$^{c}$ &  supported in part by the MINERVA Gesellschaft f\"ur Forschung GmbH, the Israel Science   
          Foundation (grant no. 293/02-11.2) and the U.S.-Israel Binational Science Foundation \\  
$^{d}$ &  supported by the German-Israeli Foundation and the Israel Science Foundation\\           
$^{e}$ &  supported by the Italian National Institute for Nuclear Physics (INFN) \\                
$^{f}$ &  supported by the Japanese Ministry of Education, Culture, Sports, Science and Technology 
          (MEXT) and its grants for Scientific Research\\                                          
$^{g}$ &  supported by the Korean Ministry of Education and Korea Science and Engineering          
          Foundation\\                                                                             
$^{h}$ &  supported by the Netherlands Foundation for Research on Matter (FOM)\\                   
$^{i}$ &  supported by the Polish State Committee for Scientific Research, project no.             
          DESY/256/2006 - 154/DES/2006/03\\                                                        
$^{j}$ &  partially supported by the German Federal Ministry for Education and Research (BMBF)\\   
$^{k}$ &  supported by RF Presidential grant N 8122.2006.2 for the leading                         
          scientific schools and by the Russian Ministry of Education and Science through its      
          grant for Scientific Research on High Energy Physics\\                                   
$^{l}$ &  supported by the Spanish Ministry of Education and Science through funds provided by     
          CICYT\\                                                                                  
$^{m}$ &  supported by the Science and Technology Facilities Council, UK\\                         
$^{n}$ &  supported by the US Department of Energy\\                                               
$^{o}$ &  supported by the US National Science Foundation. Any opinion,                            
findings and conclusions or recommendations expressed in this material                             
are those of the authors and do not necessarily reflect the views of the                           
National Science Foundation.\\                                                                     
$^{p}$ &  supported by the Polish Ministry of Science and Higher Education                         
as a scientific project (2006-2008)\\                                                              
$^{q}$ &  supported by FNRS and its associated funds (IISN and FRIA) and by an Inter-University    
          Attraction Poles Programme subsidised by the Belgian Federal Science Policy Office\\     
$^{r}$ &  supported by the Malaysian Ministry of Science, Technology and                           
Innovation/Akademi Sains Malaysia grant SAGA 66-02-03-0048\\                                       
\end{tabular}                                                                                      
                                                           %
                                                           %

\pagenumbering{arabic}
\pagestyle{plain}
\section{Introduction}
\label{introduction}

The production of multi-hadronic final states in high-energy two-body collisions
has long been a subject of great interest from experimental and theoretical points of
view. The charged multiplicity at HERA has been measured previously by the
H1~\cite{zfp:c72:573,np:b504:3,epj:c5:439,pl:b654:148} and ZEUS~\cite{zfp:c67:93,epj:c11:251,pl:b510:36}
experiments. In this paper, new measurements by the ZEUS collaboration
of the charged multiplicity in deep inelastic scattering (DIS) are presented. 

Measurements are performed in the hadronic centre-of-mass (HCM)
frame and the results are compared with those obtained in $\epem$ 
collisions, as well as with those from previous DIS experiments
~\cite{zfp:c72:573,zfp:c76:441,zfp:c35:335,zfp:c54:45}. 
For the $ep$ final state, differences are expected in the
photon (current) and proton (target) fragmentation regions, due to the asymmetric nature
of the reaction. The detector acceptance only 
allows the current fragmentation region to be measured.

Measurements of the charged multiplicity are also performed in the current 
region of the Breit frame, which should behave similarly to one hemisphere in
$\epem$ collisions. Previous DIS results ~\cite{pl:b445:439,zfp:c67:93,np:b504:3} 
using $Q$, the virtuality of the exchanged photon,
as the scale showed a reasonable agreement with $\epem$ data. 
However, this agreement degraded at
values of $Q$ below 6 -- 8 GeV. In this paper, the energy of the current
region of the Breit frame is used as the scale to
compare with $\epem$ data.

An alternative energy scale, the invariant mass of
the hadronic system, has also been used in both the Breit and HCM frames. 
The results using this variable are also compared to results from
$\epem$ collisions.

\section{Experimental set-up}

The data were collected with the ZEUS detector
during the 1996 and 1997 running periods, when HERA operated with protons of energy
$E_p=820$~GeV and positrons of energy $E_e=27.5$~GeV, and correspond to an integrated luminosity of
$38.6\pm 0.6$~pb$^{-1}$.

The ZEUS detector is described in detail elsewhere~\cite{zeus:1993:bluebook}. The most important components
used in the current analysis were the central tracking detector and the
calorimeter.

\Zctddesc\ZcoosysfnB

\Zcaldesc

\section{Data selection}

Deep inelastic scattering events were selected by requiring that the outgoing positron was measured
in the CAL. The scattered-positron identification was based on a neural-network algorithm using the CAL
information~\cite{nim:a365:508}.

For the reconstruction of the photon virtuality, $Q^2$, Bjorken $x$, and the $\gamma^*P$ centre-of-mass energy, $W$,
the double angle (DA) method was chosen, in which the scattered-positron angle,
${\theta}_e$, and the hadronic angle ${\gamma}_H$ are used~\cite{proc:hera:1991:23}. In the naive
quark-parton model, ${\gamma}_H$ is the angle of the scattered massless quark in the laboratory frame.

For each event, the measurement of the 
charged multiplicity was performed using tracks reconstructed in 
the CTD. The energy of the
hadronic final state was measured using
a combination of track and 
CAL information,
excluding the cells and the track associated with the scattered positron. 
The selected tracks and CAL
clusters were treated as massless Energy Flow Objects (EFOs)
\cite{briskin:phd:1998}. 
The clustering
of objects was done according to the Snowmass
convention\cite{proc:snowmass:1990:134}.
The transverse momentum, $p_{T}$, of each EFO was required to be greater than $0.15$~GeV.

The event selection criteria were:

\begin{itemize}
\item $E_{e^\prime}>12$~GeV, where $E_{e^\prime}$ is the energy of the scattered
      positron, to select neutral current DIS events;
\item $y_e\,\leq\,0.95$, where $y_e$ is the scaling variable $y$ as determined from the energy
      and polar angle of the scattered positron. This cut reduces the photoproduction background;
\item $y_{\rm JB}\,\geq\,0.04$, where $y_{\rm JB}$ is the estimate of $y$ using
      the Jacquet-Blondel (JB) method~~\cite{proc:epfac:1979:391}. It is defined as 
      $y_{\rm JB}= \sum_{h}(E_h-P_{Z_h})/2E_e$, where the sum 
      runs over all EFOs and $E_h$ and $P_{Z_h}$ are the energies and longitudinal 
      momenta of the EFOs respectively. This requirement guarantees sufficient 
      accuracy for the DA reconstruction method;
\item $35\,\leq\,\delta\,\leq\,60\,\Gev$, where $\delta=\sum_{i}(E_i-P_{Z_i})$\ and the sum 
      runs over all EFOs and the scattered positron, to remove photoproduction
      events and events with large  radiative corrections;
\item $R \ge 25\, \cm$, where $R$ is the distance from the beam axis to the
      impact position of the scattered positron on the face of the CAL.
      This ensured that the positron was fully contained within the detector and
      its position reconstructed with sufficient accuracy;
\item $|Z_{\rm vtx}|<50\,\cm$, where  $Z_{\rm vtx}$ is the longitudinal position of the
      vertex, to reduce background events from non-$ep$ collisions.
\end{itemize}

To ensure high-quality tracks reconstructed with high
efficiency, the following requirements were made:

\begin{itemize} 
\item the tracks had to pass through at least three CTD superlayers;
\item the tracks had to be associated with the primary event vertex; 
\item the tracks were restricted to the region $|\eta_{\rm LAB}| \le 1.75$, where
      $\eta_{\rm LAB}$
      is the pseudorapidity of the tracks in the laboratory frame;
\item the tracks had $p_{T}>0.15\,\gev$.
\end{itemize}

The analysis was restricted to the kinematic 
range $Q^2>25$~GeV$^2$ and $70<W< 225$~GeV.

\section{Analysis method}
\label{analysis}

Due to the large asymmetry of the beam 
energies at HERA, a large fraction of the hadronic final state close to the proton
direction lies outside the detector acceptance.
Therefore
only hadrons belonging to the 
current fragmentation regions of the HCM 
and Breit frames were used in this analysis.
The boost to the corresponding reference frames was calculated using the 
positron four-momentum taken from the DA method.

\subsection{Breit frame}
\label{Breit}

In the Breit frame, which is defined by the condition that the momentum of
the exchanged virtual boson is purely spacelike, ${\bf q} = (0,0,-Q)$, 
the particles produced in the interaction can be
assigned to one of two regions: the current region if their longitudinal momentum in the Breit
frame is negative, and the target region if their longitudinal momentum is positive. 
The hadronic system of the current region used in this analysis
is almost fully (about 95$\%$) contained within the acceptance of the CTD.

Previous analyses compared the mean charged multiplicity $\nch$ as a function
of $Q$ to $\nch /2$ as a function of $\sqrt{s}$ in $\epem$ 
collisions~\cite{np:b504:3,zfp:c67:93}. 
For values of $Q>8$~GeV, reasonable agreement was observed, while some
disagreement was found for $Q<8$~GeV. 
The difference can be understood in terms of higher-order
processes~\cite{zfp:c2:237}, which change the available energy in the current
region of the Breit frame, $\ecrbreit$, which is no longer
equal to $Q/2$.
In this analysis the quantity $\twoecrbreit$ is used as a scale. 
On an event-by-event basis this method should compensate for
particles and their corresponding energies 
migrating  
between current and target regions of the Breit frame.

\subsection{Hadronic centre-of-mass frame}
\label{HCM}

In the HCM frame, the exchanged virtual boson has four-momentum 
${\bf q} = {\{E,p\}} = (\frac{W^2-Q^2}{2W},0,0,\frac{W^2+Q^2}{2W})$.
The hadronic final state is separated into the photon (current)
and proton (target) fragmentation regions.
About 60 -- 80\% of the current region of the HCM frame  
is contained within the acceptance of the CTD.
 
The multiplicity in the HCM frame in DIS is usually studied as a function of 
$W$~\cite{zfp:c72:573,zfp:c76:441,zfp:c35:335,zfp:c54:45}. At
HERA,  
the energy in the current region of the HCM frame, $E_{\rm HCM}^{\rm cr}$,
 coincides with $W/2$ within $0.3-0.4\%$.
Thus, for practical reasons, $W$ was used as the energy scale.

\subsection{Invariant mass of the hadronic system}
\label{Meff}

Charged multiplicities in the current region of both the Breit
and HCM frames were also measured as functions of
the invariant mass:

\begin{equation} M^2_{\rm eff}=\left(\sum_{i}E_i \right)^2-\left( \sum_{i} P_{X_i} \right)^2 - 
\left( \sum_{i} P_{Y_i} \right)^2 -
\left( \sum_{i} P_{Z_i} \right)^2 ,\end{equation}

where the sum runs over all charged and neutral 
particles of the corresponding hadronic system.

\section{Monte Carlo models, acceptance corrections and systematic errors}
\label{MC_correction}

Samples of neutral current DIS events were generated using 
the colour-dipole model
as implemented in {\sc Ariadne}~4.12~\cite{cpc:71:15} or with 
the MEPS model of {\sc Lepto~}6.5~\cite{cpc:101:108}.
Both programs were interfaced to
 {\sc Heracles~}4.6.1~\cite{cpc:69:155,*spi:www:heracles}
using the {\sc Djangoh~}1.1~\cite{spi:www:djangoh11,*cpc:81:381} program.
Both {\sc Ariadne} and {\sc Lepto} use the Lund string model~\cite{prep:97:31} 
for the hadronisation. The event samples were generated using the CTEQ4D~\cite{pr:d55:1280}
parameterisation of the parton distribution function (PDF) in the proton.
The cluster hadronisation
model as implemented in {\sc Herwig 6.1}~\cite{cpc:67:465} was used to 
estimate the effect of different hadronisation schemes 
on the unfolding procedure. The minimum transverse momentum of outgoing
partons in the hard interaction and of partons participating in multi-parton
interactions is adjustable in {\sc Herwig} using the parameter $p_{T}^{\rm min}$.
The parameter was tuned in this analysis to improve the description of the detector 
distributions. The best agreement
was found for $p_{T}^{\rm min}=2.5$~GeV~\cite{epj:c50:283}. The HERWIG samples were generated 
using CTEQ5L~\cite{epj:c12:375} proton PDF parametrisation. The MC samples were
used both for data correction and for comparison of the data to the model
predictions.

The corrections applied to the data accounted for the effects of acceptance and 
resolution of the detector,
event selection cuts, QED-radiative effects, track reconstruction, 
track selection cuts, 
and energy losses in the inactive
material in front of the calorimeter in the case of the energy measurement.
Finally the multiplicity distributions were corrected 
using a matrix unfolding method as described in earlier studies~\cite{zfp:c67:93}.

The generated events were passed through a full simulation of
the detector, using a program based on {\sc Geant}~3.13~\cite{geant}, and
processed and selected with the same programs as used for the data.
The simulated samples were used to determine the response of the 
detector and to 
evaluate the correction factors necessary to obtain hadron-level quantities. 
The hadron level is defined by those hadrons with lifetime 
$\tau \ge 3\cdot10^{-10}$~seconds. 
In order to compare the results to different experiments, 
corrections were calculated both including and not including the decay 
products of 
$K^0_S$ and $\Lambda$.

The dominant systematic  
uncertainty in this analysis arises from the simulation of the hadronic final state.
To correct the data, 
the average of the correction factors from the 
{\sc Ariadne} and {\sc Herwig} MC programs was used.
One half of the difference, as large as 
 $5\%$,
was assigned to the systematic uncertainties.

Other sources of uncertainty are (typical
values of the uncertainties are shown in parentheses):
event reconstruction and
selection ($<0.5\%$), track reconstruction and selection ($<0.5\% $), and
the uncertainty due to variation of the $Q^2$ cut by its resolution ($<1.7\%$).
The uncertainty due to contamination from diffractive events is negligible.
The individual systematic uncertainties were added in quadrature.
The major correlated uncertainty comes from the CAL energy scale ($<1.5\% $)
and is not shown in tables and figures.
A detailed study of the sources contributing to the uncertainties of the
measurement can be found in~\cite{rosin:phd:2006}.
\section{Results}

\subsection{Multiplicity distributions}

\begin{figure}
\centerline{\epsfig{file=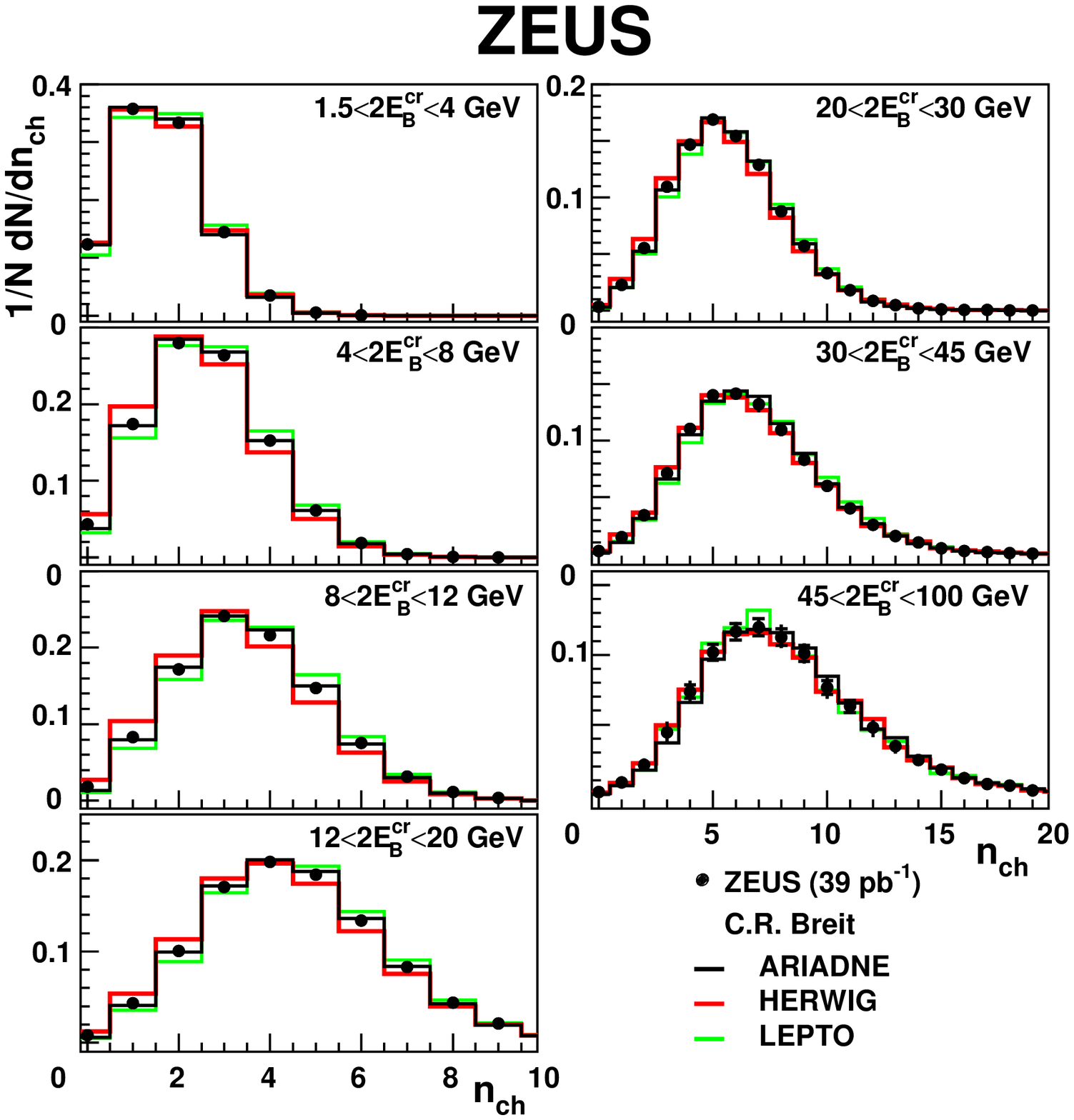,width=\textwidth,clip=}}
\caption{Multiplicity distributions measured in the current region of the Breit
frame in bins of $\twoecrbreit$ (solid circles). 
The inner error bars represent the statistical uncertainties and the outer error bars
the statistical and systematic uncertainties added in quadrature. 
The predictions (solid lines)
of different MC models are also shown.
}
\label{fig1}
\end{figure}

\begin{figure}
\centerline{\epsfig{file=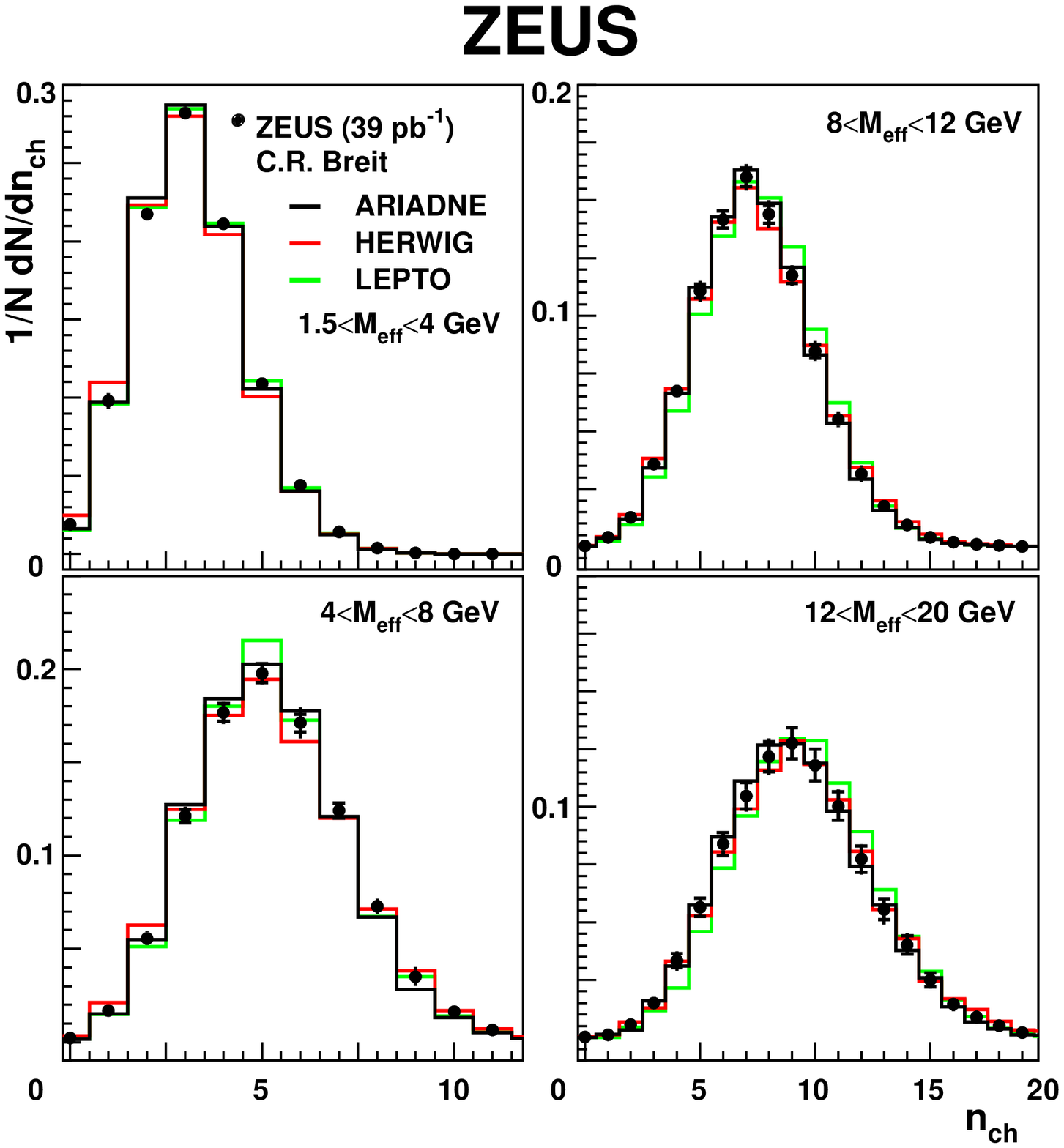,width=\textwidth,clip=}}
\caption{Multiplicity distributions measured in the current region of the Breit
frame in bins of $\meff$ (solid circles). 
The inner error bars represent the statistical uncertainties and the outer error bars
the statistical and systematic uncertainties added in quadrature. 
The predictions (solid lines) 
of different MC models are also shown.
}
\label{fig2}
\end{figure}

The multiplicity distributions in the current region of the Breit frame
are presented in Fig.~\ref{fig1} and Table~\ref{dist_2ebr} and 
in Fig.~\ref{fig2} and Table~\ref{dist_brmeff} 
in bins of $\twoecrbreit$ and in bins of $\meff$ respectively. 
The kinematic range of the analysis restricts the 
$\meff$ measurements in the current region of the Breit frame to a maximum value 
of about 20~GeV. 
The predictions of {\sc Ariadne}, {\sc Lepto} and {\sc Herwig} are also shown. 
All three MC models generally describe the data, but {\sc Ariadne} 
gives the best description. In all Figures and Tables the 
charged-particle decay products of $K^0_S$ and $\Lambda$ are included, unless
otherwise stated.

For a given bin of energy, the multiplicities as a function of $\twoecrbreit$ and
$\meff$ differ by approximately a factor of two. This is due to
the fact that $\twoecrbreit$ (as well as $Q$ or $W$) characterise the
total centre-of-mass energy of the system of which only one hemisphere 
is measured. On the other hand the $\meff$ method measures the multiplicity of the 
system with respect to the corresponding invariant mass.

\begin{figure}
\centerline{\epsfig{file=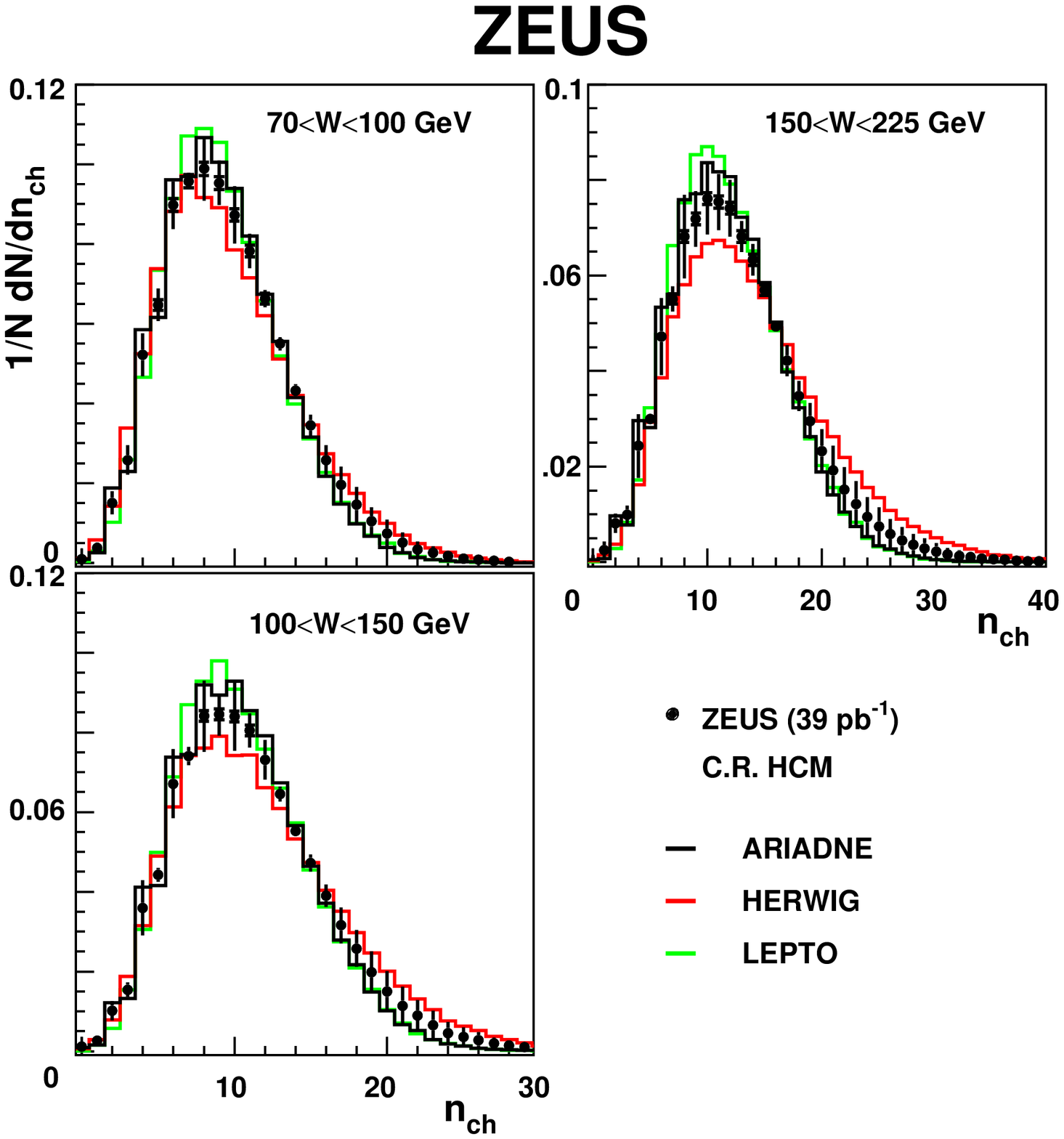,width=\textwidth,clip=}}
\caption{Multiplicity distributions measured in the current region of the HCM
frame in bins of $W$ (solid circles). 
The inner error bars represent the statistical uncertainties and the outer error bars
the statistical and systematic uncertainties added in quadrature. 
The predictions (solid lines) 
of different MC models are also shown.
}
\label{fig3}
\end{figure}

The multiplicity distributions in the current region of the HCM frame are
presented in Fig.~\ref{fig3} and Table~\ref{dist_whcm} 
in bins of $W$.
Both {\sc Ariadne} and {\sc Lepto} predict similar $W$ 
distributions and give a reasonable description of the data. {\sc Herwig}
predicts longer tails for the multiplicity distributions in bins of $W$. This 
leads to 
higher average multiplicities, 
affecting the unfolded multiplicity values and increasing the systematic uncertainties
of the measurement.  
The {\sc Ariadne} predictions vary slightly from both {\sc Lepto} and {\sc Herwig} 
at the rising edge of the distributions, which 
also leads to an increase of the systematic uncertainties.

\begin{figure}
\centerline{\epsfig{file=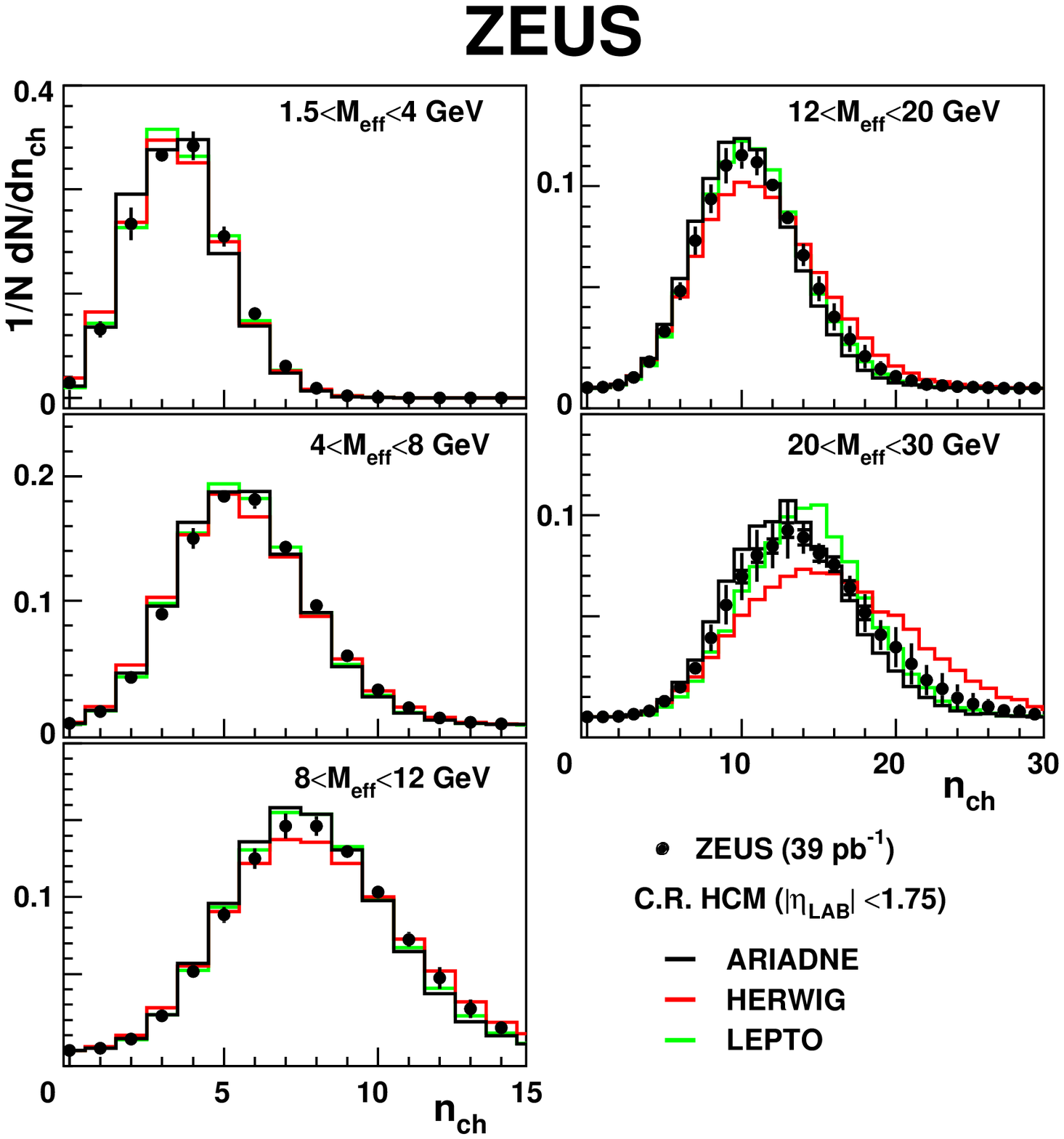,width=\textwidth,clip=}}
\caption{Multiplicity distributions measured in the current region of the HCM
frame in bins of $\meff$ (solid circles). 
The inner error bars represent the statistical uncertainties and the outer error bars
the statistical and systematic uncertainties added in quadrature. 
The predictions (solid lines) 
of different MC models are also shown.
}
\label{fig4}
\end{figure}

The multiplicity distributions in the current region of the HCM frame 
in bins of $\meff$ are presented in Fig.~\ref{fig4} and Table~\ref{dist_hcmmeff}. 
To minimise the extrapolation both in multiplicity and $\meff$, an additional 
requirement, $|\eta_{\rm LAB}| < 1.75$, was applied at the hadron level.
None of the MC models shown in Fig.~\ref{fig4} give a complete description of 
the data. This is most visible at higher values of $\meff$.
  
\subsection{KNO scaling}

The multiplicity distributions are expected to scale with energy as
discussed in detail elsewhere~\cite{mpl:a6:981}. 
A commonly used form of the scaling, from KNO~\cite{np:b40:317}, is shown 
in Figs.~5--8, where the product of 
the multiplicity distribution $P(n_{\rm ch})$ with average multiplicity $\nch$, $\nch P(n_{\rm ch})$,
is shown as a function of $n_{\rm ch}/ \nch $.

\begin{figure}
\begin{center}
\psfigadd{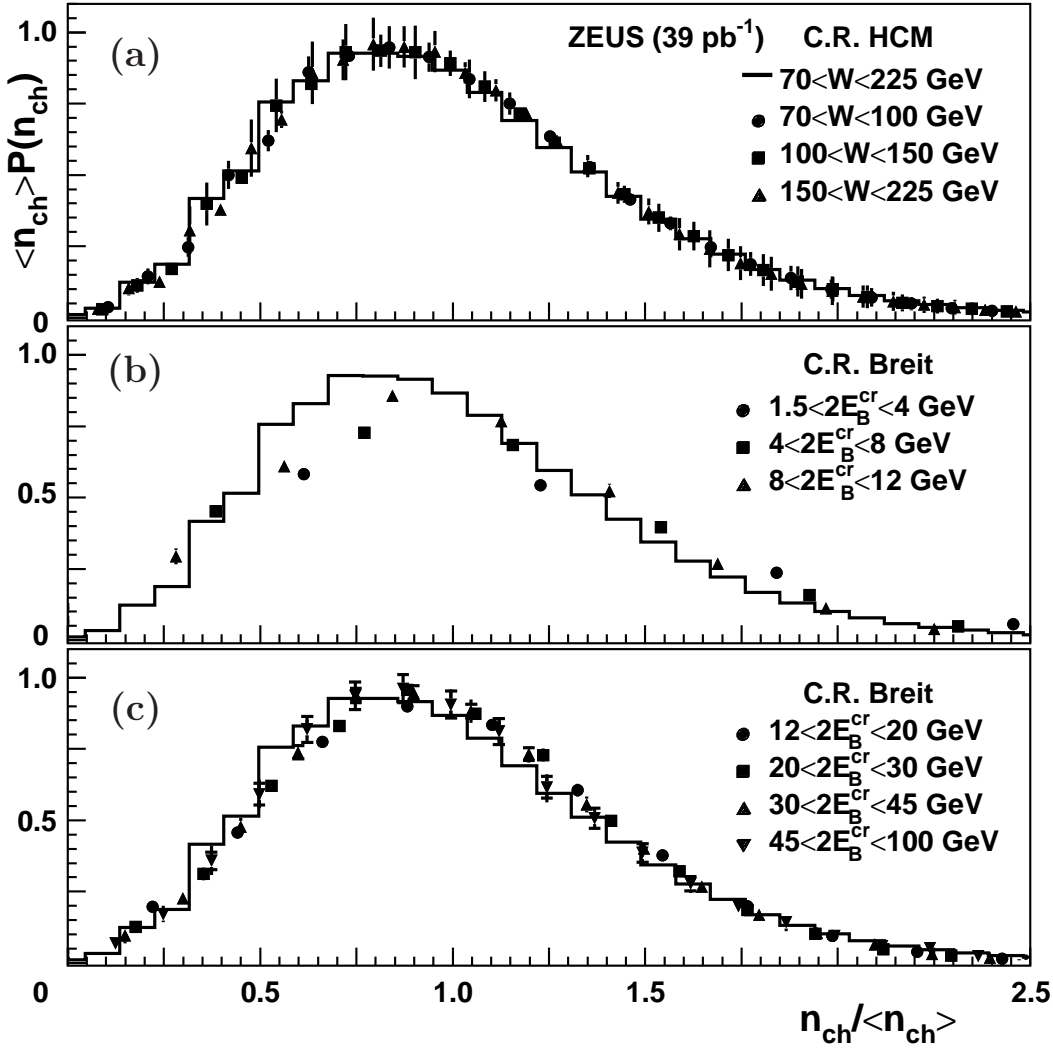}{\textwidth}{\textwidth}{%
                   \Text(250,1370)[]{\large\bf (a)}
                   \Text(250,950)[]{\large\bf (b)}
                   \Text(250,500)[]{\large\bf (c)}}
\end{center}
\caption{ The multiplicity distributions plotted in KNO form. 
The reference KNO histogram represents 
the measured distribution in the HCM frame in the entire $W$ range.
(a) Multiplicity distributions measured in the current region of the HCM
frame in bins of $W$. 
Multiplicity distributions measured in the current region of the Breit
frame in bins of $\twoecrbreit$ for (b) $1.5 < \twoecrbreit < 12~{\rm GeV}$
and (c) $12 < \twoecrbreit < 100~{\rm GeV}$. 
}
\label{fig5}
\end{figure}

\begin{figure}

\vspace*{3cm}
\begin{center}
\psfigadd{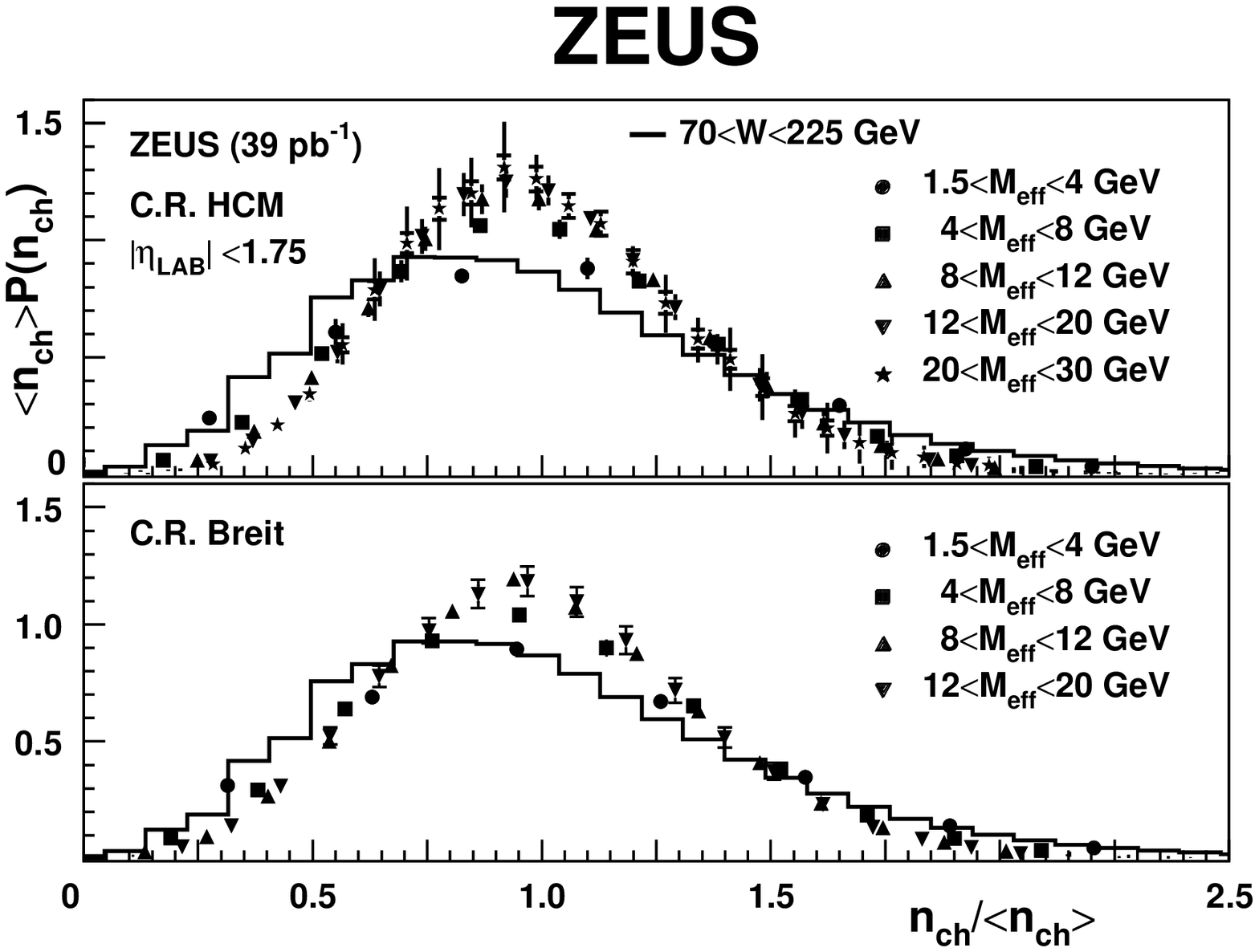}{\textwidth}{\textwidth}{%
                   \Text(250,1170)[]{\large\bf (a)}
                   \Text(250,750)[]{\large\bf (b)}}
\end{center}

\vspace*{-3cm}
\caption{The multiplicity distributions plotted in KNO form and compared to the
reference KNO distribution (histogram). 
The multiplicity distributions are measured in bins of $\meff$
in the current regions of (a) the HCM
frame, restricted in $\eta_{\rm LAB}$, and 
(b) the Breit frame.
}
\label{fig6}
\end{figure}

In Fig.~\ref{fig5}(a), the KNO distributions measured in bins
of $W$ in the current region of the HCM frame are shown. 
Within the uncertainties,
the distributions measured in three bins of $W$ agree. They also
agree well with the average distribution, which was calculated using data for  
the entire $W$ region, $70 < W < 225$~GeV. This average KNO spectrum, presented as
a histogram, is shown in Figs.~\ref{fig5} and~\ref{fig6} as a reference KNO distribution.

The reference KNO distribution is compared to the measurements 
in the current region of the Breit frame in 
Figs.~\ref{fig5}(b) and \ref{fig5}(c) in bins of $\twoecrbreit$. For values of 
$\twoecrbreit > 12$~GeV, as shown in Fig.~\ref{fig5}(c), the measurements
are in reasonable agreement with the reference KNO histogram. For lower values of $\twoecrbreit$,
as demonstrated in Fig.~\ref{fig5}(b), the distributions do
not follow the KNO-scaling behaviour; they have different shapes, but
approach the KNO curve with increasing values of $\twoecrbreit$.

In Fig.~\ref{fig6}, the KNO distributions measured in bins
of $\meff$ in the current region of the HCM frame are presented. 
The multiplicity distributions in bins of $\meff$ do not follow the
same KNO scaling observed for measurements as functions of $W$ or $\twoecrbreit$, 
but do demonstrate scaling behaviour for $\meff$
values above 8~GeV. The measurements at $\meff < 4$~GeV in the current regions of both the 
Breit and HCM frames 
behave differently from the measurements at the higher values of $\meff$.

\begin{figure}
\begin{center}
\psfigadd{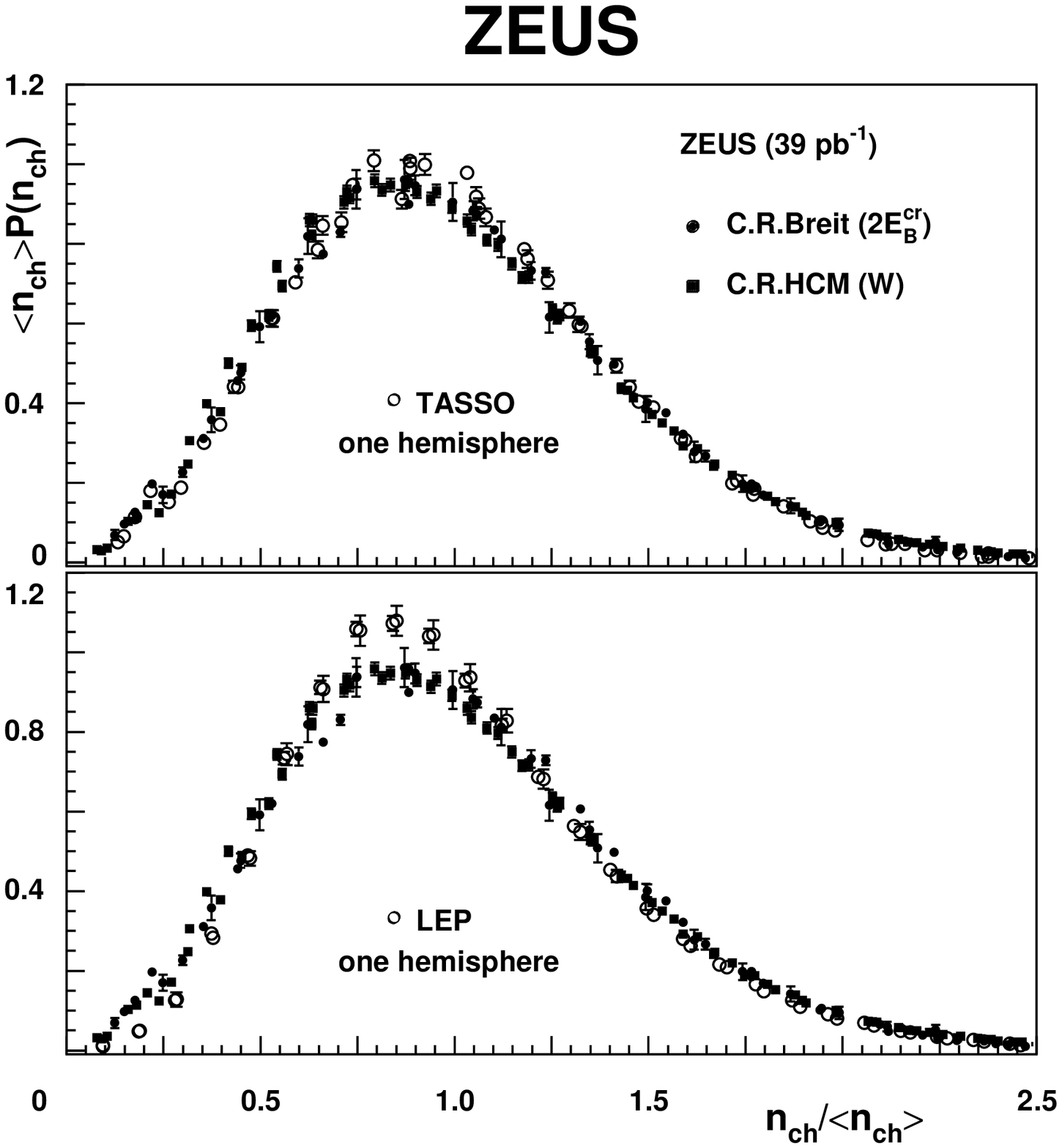}{\textwidth}{\textwidth}{%
                   \Text(250,1370)[]{\large\bf (a)}
                   \Text(250,700)[]{\large\bf (b)}}
\end{center}
\caption{The multiplicity distributions plotted in KNO form and compared to the
results of $\epem$ experiments. The solid circles represent the ZEUS data measured
in the current region of the Breit frame in bins of $\twoecrbreit$, for 
$\twoecrbreit \ge 12~{\rm GeV}$, and 
the solid squares represent the data measured in the current region of the 
HCM frame in bins of $W$.
Multiplicities measured in one hemisphere of the $\epem$ collision are shown in
bins of $\sqrt{s_{ee}}$ for (a) the TASSO experiment ~\protect\cite{zfp:c45:193},
 in the energy range $14 \le \sqrt{s_{ee}} \le 44~{\rm GeV}$, and 
(b) for the LEP experiments~\protect\cite{zfp:c35:539,zfp:c50:185}, measured at energy
$\sqrt{s_{ee}} = 91.2~{\rm GeV}$.
}
\label{fig6_2}
\end{figure}

Figure ~\ref{fig6_2} shows a comparison of the 
KNO distributions from ZEUS
with results obtained in $\epem$ collisions.
The measurements in bins
of $\twoecrbreit$, for $\twoecrbreit > 12$~GeV, and in bins of $W$
are plotted together and compared with measurements in one hemisphere
of $\epem$. 
In Fig.~\ref{fig6_2}(a), a comparison with results from     
 the TASSO collaboration~\cite{zfp:c45:193} in the energy range
$14 < \sqrt{s_{ee}} < 44$~GeV is shown.
At LEP only DELPHI~\cite{zfp:c50:185} and OPAL~\cite{zfp:c35:539} performed 
measurements in a single hemisphere at $\sqrt{s_{ee}} = 91.2$~GeV.  
A comparison with the present data is shown in 
Fig.~\ref{fig6_2}(b). 
The systematic uncertainties are not shown, 
but within statistical  uncertainties there is a remarkable 
 agreement between $ep$ and $\epem$ results. 
However, the LEP data differ somewhat 
from the present measurement
in the peak region and at very low values of $n_{\rm ch}/ \nch $.

\begin{figure}
\begin{center}
\psfigadd{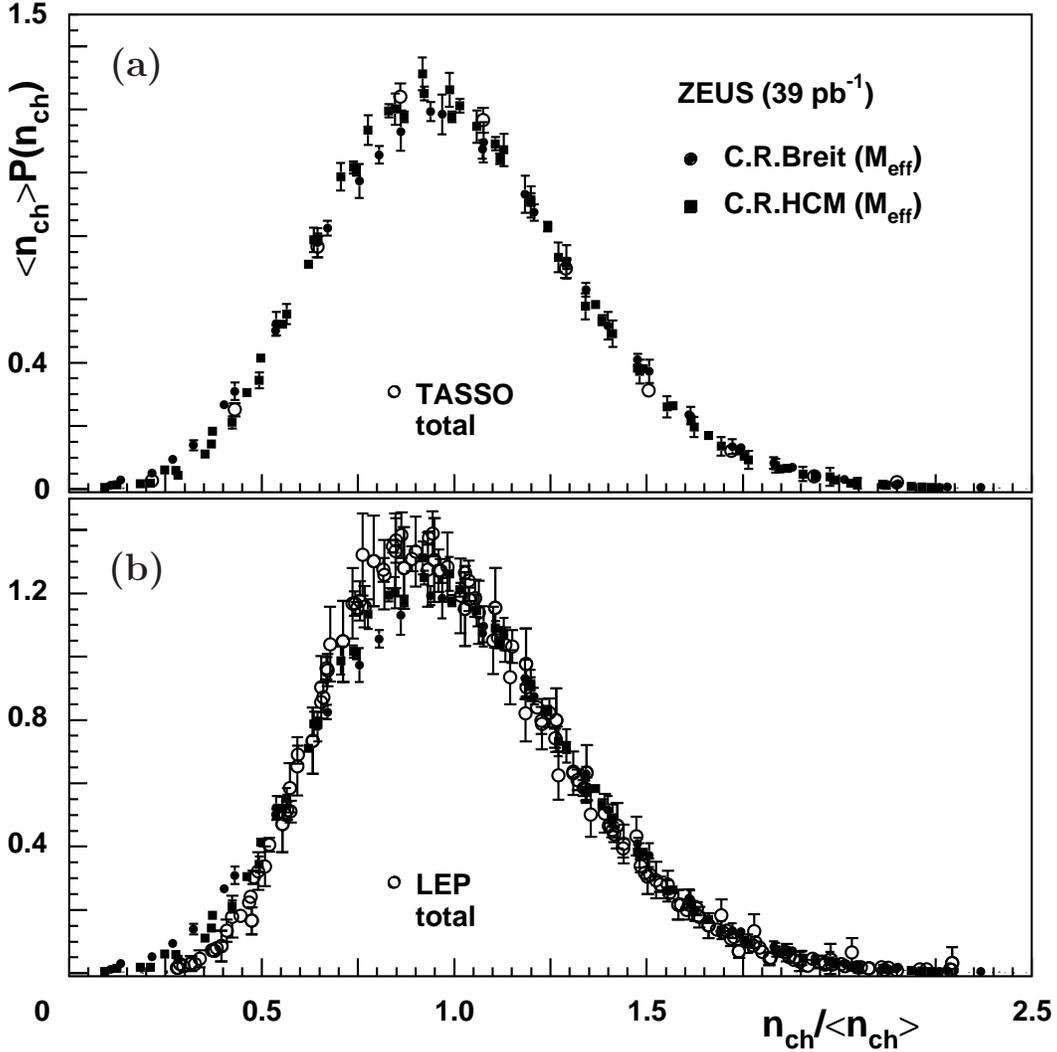}{\textwidth}{\textwidth}{%
                   \Text(250,1370)[]{\large\bf (a)}
                   \Text(250,700)[]{\large\bf (b)}}
\end{center}
\caption{The multiplicity distributions plotted in KNO form and compared to the
results of the $\epem$ experiments. The solid circles represent the ZEUS data measured
in the current region of the Breit frame and
the solid squares represent the data measured in the current region of the 
HCM 
frame both in bins of 
$\meff$, for $\meff \ge 8~{\rm GeV}$.
Charged multiplicities measured for the whole event in $\epem$ collisions are shown in
bins of $\sqrt{s_{ee}}$ for (a) the TASSO experiment ~\protect\cite{zfp:c45:193},
 in the energy range $14 \le \sqrt{s_{ee}} \le 44~{\rm GeV}$, and 
(b) for the LEP experiments~\protect\cite{zfp:c35:539,zfp:c50:185,pl:b273:181,*zfp:c73:409,
*pl:b372:172,*pl:b416:233,*epj:c18:203,*pr:399:71,*pl:b577:109,
*pl:b371:137,*pl:b404:390,*pl:b444:569,*cern-ppe/96-47,*cern-ppe/97-015,
*cern-ep/99-178} in the energy range
$91.2 \le \sqrt{s_{ee}} \le 209~{\rm GeV}$.
}
\label{fig6_3}
\end{figure}

The data as a function of $\meff$, for $\meff > 8$~GeV, are compared with the 
$\epem$ measurements for the whole event in Fig.~\ref{fig6_3}.
Both the TASSO  
 and LEP~\cite{zfp:c50:185,zfp:c35:539,pl:b273:181,*zfp:c73:409,
*pl:b372:172,*pl:b416:233,*epj:c18:203,*pr:399:71,*pl:b577:109,
*pl:b371:137,*pl:b404:390,*pl:b444:569,*cern-ppe/96-47,*cern-ppe/97-015,
*cern-ep/99-178} data
 ($91.2< \sqrt{s_{ee}} < 209$~GeV) 
 agree 
 with the present measurement.

\subsection{Mean charged multiplicity}
\label{means}

\begin{figure}
\centerline{\epsfig{file=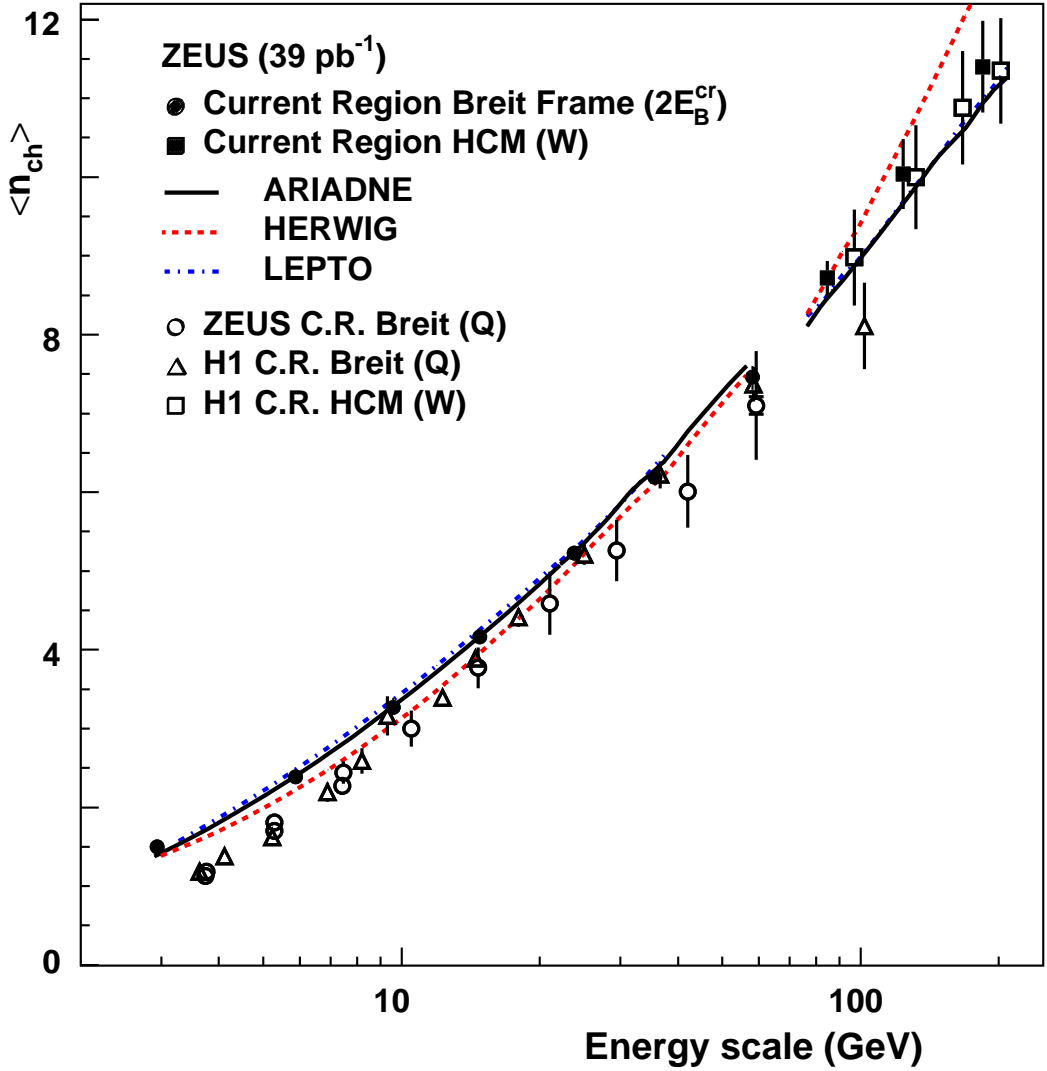,width=\textwidth,clip=}}
\caption{
Mean charged multiplicity, $\langle n_{\rm ch}\rangle$, in the 
current region of the Breit frame as a function of $\twoecrbreit$ and 
in the current
fragmentation region of the HCM frame as a function of $W$.
The inner error bars
represent the statistical uncertainties, typically smaller than the size of the symbols. 
The outer error bars
represent the quadratic sum of statistical and systematic uncertainties.
Also shown are the results of previous HERA measurements 
~\protect\cite{pl:b654:148,zfp:c72:573,np:b504:3,zfp:c67:93} and 
predictions from {\sc Ariadne}, {\sc Lepto} and {\sc Herwig}.
The decay products of $K^0_S$ and $\Lambda$ are not included.
}
\label{fig7}
\end{figure}

Figure~\ref{fig7} and Table~\ref{ave_2e} show the mean charged multiplicity, $\nch$,
in the current region of the HCM frame as a function of $W$
and the mean charged multiplicity in the current region of the Breit frame
as a function of $\twoecrbreit$. 
The $K^0_S$ and $\Lambda$ hadrons were considered stable in Fig.~\ref{fig7}, 
where the data are
compared with results of previously published HERA measurements
~\cite{pl:b654:148,zfp:c72:573,np:b504:3,zfp:c67:93}.  
As expected, at low values of $\twoecrbreit$, the
measurement differs with those as a function of $Q$
(see Section~\ref{Breit}). 
At higher values of $\twoecrbreit$ the
data agree within the experimental uncertainties with the previous 
ZEUS and  H1 measurements,
but lie systematically above them. The data are in good agreement with the 
{\sc Ariadne} and {\sc Lepto} predictions. The {\sc Herwig} predictions also describe 
 the data but are below those from {\sc Ariadne} and {\sc Lepto}.
In the current region of the HCM, the measurement agrees well,
with improved statistical and systematic uncertainties, 
with the earlier H1 results.
The {\sc Ariadne} and  {\sc Lepto} predictions agree with the data. 
{\sc Herwig} predicts 
a very different slope, with much higher multiplicities at higher energies;  
with increasing energy the agreement with data degrades.

\begin{figure}
\begin{center}
\psfigadd{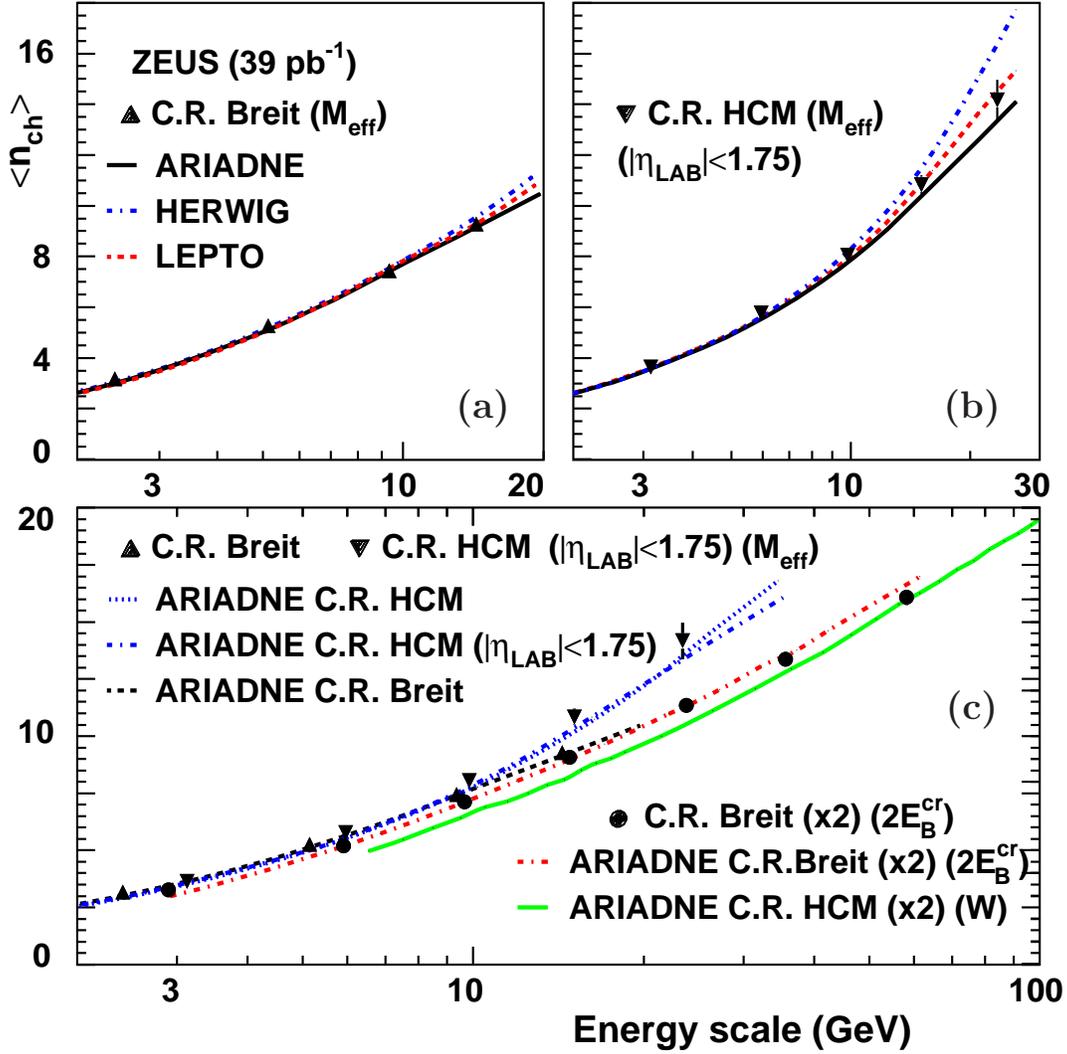}{\textwidth}{\textwidth}{%
                   \Text(700,900)[]{\large\bf (a)}                   
		   \Text(1350,900)[]{\large\bf (b)}
                   \Text(1350,500)[]{\large\bf (c)}}
\end{center}
\caption{
Mean charged multiplicity, $\nch$, measured as a function of $\meff$ 
(a) in the current region of the Breit frame and  
(b) in the current region of the HCM frame compared to MC predictions.
(c) Comparison between measurements in the current regions of the Breit 
and HCM frame as functions of $\meff$ and with the measurement
as a function of $\twoecrbreit$. 
The predictions from {\sc Ariadne} are also shown. 
}
\label{fig8}
\end{figure}

The mean
charged multiplicities in the current regions of the Breit and HCM frames  
are presented 
in Fig.~\ref{fig8} and Table~\ref{ave_meff} 
as a function of the invariant mass of the corresponding hadronic
system, $\meff$. 
In Figs.~\ref{fig8}(a) and~\ref{fig8}(b), the multiplicities are compared to the MC predictions.
All three MC models describe the data 
reasonably well; however in the last $\meff$ bin
in the current region of the HCM, the {\sc Herwig} prediction is too
high.
In Fig.~\ref{fig8}(c), both measurements are shown together and compared with different 
MC curves calculated using the {\sc Ariadne} MC.
The measurements in the Breit and HCM frames agree at values of 
$\meff$ less than 10 GeV. Above this value, $\nch$ rises much 
faster with $\meff$ in the current region of the HCM frame than in the 
current region of the Breit frame. Since the HCM measurement was restricted in
$\eta$, a separate {\sc Ariadne} calculation was performed 
in the total current region of the HCM frame. The difference 
is small, although the rise  
of $\nch$ with $\meff$ is faster in the total current region of the HCM frame.

Figure~\ref{fig8}(c) also shows 
$\twonch$ as a function of $\twoecrbreit$. 
This measurement exhibits the same behaviour as 
$\nch$ as a function of $\meff$ in the current region of the Breit
frame but differs from that 
 in the HCM frame.
The multiplicity in the current region of the HCM frame 
rises much faster with the invariant mass than with $W$.

\begin{figure}
\centerline{\epsfig{file=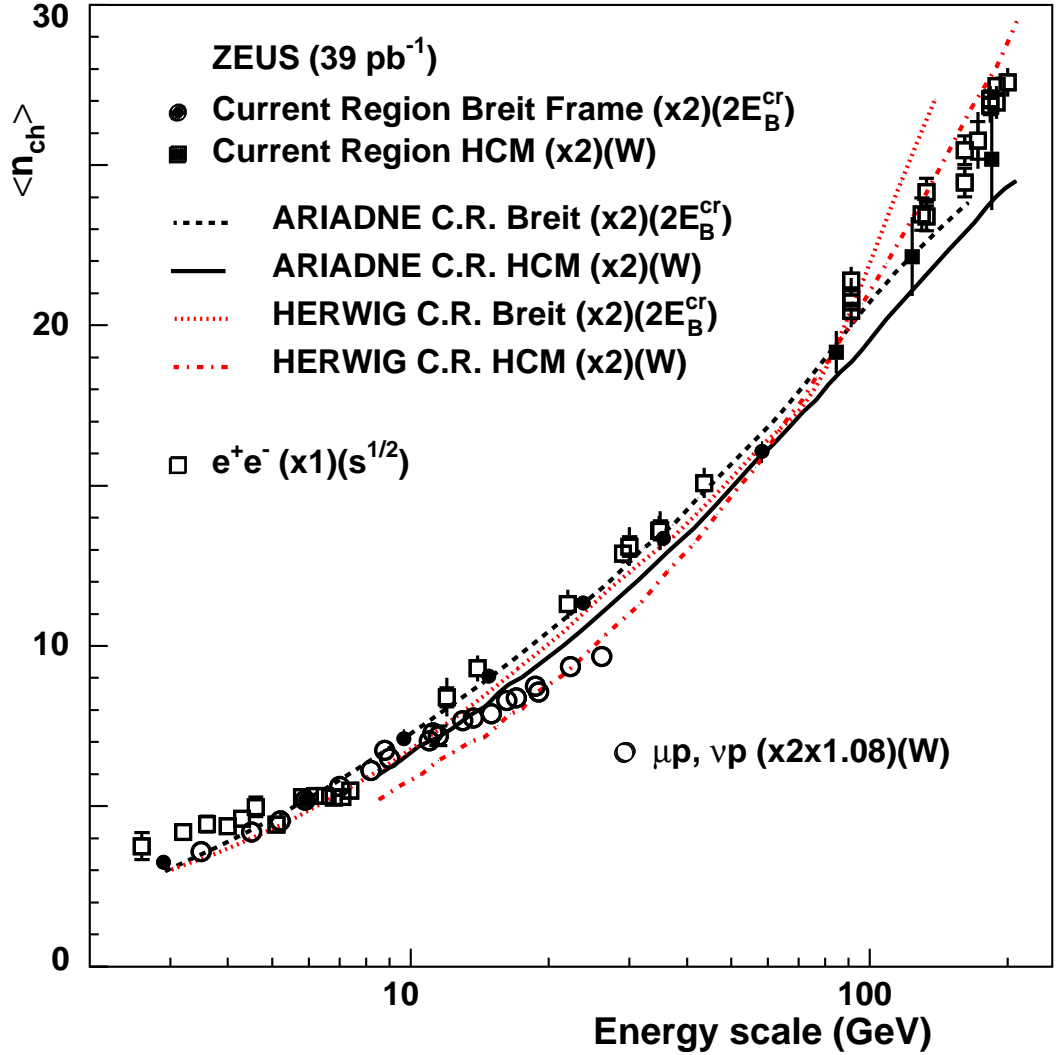,width=\textwidth,clip=}}
\caption{
Mean charged multiplicity, $\nch$, in the current region
of the Breit frame multiplied by 2 as a function of $\twoecrbreit$
and in the current region of the HCM frame multiplied by 2 as a function of $W$.
The results of 
$\epem$~\protect\cite{pl:b70:120,*zfp:c20:187,*pr:d34:3304,
zfp:c45:193,zfp:c35:539,zfp:c50:185,pl:b273:181,*zfp:c73:409,
*pl:b372:172,*pl:b416:233,*epj:c18:203,*pr:399:71,*pl:b577:109,
*pl:b371:137,*pl:b404:390,*pl:b444:569,*cern-ppe/96-47,*cern-ppe/97-015,
*cern-ep/99-178}
and fixed-target DIS 
experiments~\protect\cite{zfp:c76:441,zfp:c35:335,zfp:c54:45}
are shown. The factor 1.08 was estimated using MC predictions to correct 
the fixed-target data for the decay products of 
$K^0_S$ and $\Lambda$. 
The predictions of {\sc Ariadne} and {\sc Herwig} are also shown.
}
\label{fig9}
\end{figure}

Finally, Fig.~\ref{fig9} combines the mean charged multiplicities measured 
in the current regions of the Breit and HCM frames as functions of the 
respective energy scales, $\twoecrbreit$ and $W$. Also shown are the measurements 
from $\epem$~\cite{pl:b70:120,*zfp:c20:187,*pr:d34:3304,zfp:c45:193,
zfp:c35:539,zfp:c50:185,pl:b273:181,*zfp:c73:409,
*pl:b372:172,*pl:b416:233,*epj:c18:203,*pr:399:71,*pl:b577:109,
*pl:b371:137,*pl:b404:390,*pl:b444:569,*cern-ppe/96-47,*cern-ppe/97-015,
*cern-ep/99-178} and fixed-target~\cite{zfp:c76:441,zfp:c35:335,zfp:c54:45} experiments. 
The fixed-target data were scaled by a factor two, since they only measure one 
hemisphere and by a factor 1.08, to correct for the decays of the $K^0_S$ and 
$\Lambda$, as estimated using {\sc Ariadne} MC.

The measurements presented in this paper show good overall agreement with those 
from other experiments, exhibiting approximately the same dependence 
of the mean charged multiplicity on the respective energy scale.
At low values of the energy, the measurement as a function 
of $\twoecrbreit$ agrees well with $\epem$ data, in contrast to the previous 
measurements as a function of $Q$~\cite{np:b504:3, zfp:c67:93}. 
The measurements in the current region of the HCM
agree with the LEP data, but are systematically below them.
The data from fixed-target DIS experiments~\cite{zfp:c72:573,zfp:c76:441,zfp:c35:335,zfp:c54:45}
deviate from the observed energy dependence at energies above 15~GeV.
The {\sc Ariadne} MC prediction generally describes the energy dependence 
of the data over the entire region. However, the prediction in the HCM frame is 
generally lower than the data and than the prediction in the Breit frame. 
The {\sc Herwig} MC model does not give a good overall description of the data.

\section{Summary and conclusions}

The charged multiplicity distributions and  the mean charged multiplicity
have been  investigated in inclusive neutral current deep inelastic
$ep$ scattering in the kinematic range $\q2 >25$~GeV$^2$ and $70<W<225$~GeV 
in terms of different energy scales. The scale $\twoecrbreit$, was used 
in the current region of the Breit frame. In the current region of the HCM frame,
$W$ was used and  the invariant mass, $\meff$, was used in both frames.

In terms of KNO scaling, the charged multiplicities in the current regions of the 
Breit and HCM frames exhibit the same
behaviour as those in one hemisphere of $\epem$ collisions when $\twoecrbreit$ or $W$ 
are considered. When the energy scale $\meff$ is used, the charged multiplicities 
exhibit the same KNO-scaling behaviour as those for the whole $\epem$ event.

The mean charged multiplicity in the current region of the Breit frame scales
with $\meff$ in the same way as $\twonch$ scales with $\twoecrbreit$ and,
therefore, as $\nch$ scales with $\sqrt{s_{ee}}$ in $\epem$ collisions. 
The mean charged multiplicity in
the current region of HCM frame as a function of $\meff$ rises faster
than that in the current region of the Breit frame.

The energy scale $\twoecrbreit$, rather than $Q$, gives better agreement between 
the mean charged multiplicity measured in the current region of the Breit frame and 
that measured in $\epem$ collisions. The measurements of $\nch$ as a function of $W$ 
agree, within the uncertainties, with the data from $\epem$ collisions. When using
these scales, $ep$ DIS data can be consistently compared with data from $\epem$, 
$\mu P$ and $\nu P$ scattering over a wide energy region.

\section*{Acknowledgements}

  We are grateful to the DESY directorate for their strong support and 
  encouragment. The effort of the HERA machine group is gratefully 
  acknowledged. We thank the DESY computing and network services for their 
  support. The design, construction and installation of the ZEUS detector 
  has been made possible by the efforts of many people not listed as 
  authors.

{
\def\bibname{\Large\bf References}
\def\refname{\Large\bf References}
\pagestyle{plain}
\ifzeusbst
  \bibliographystyle{./BiBTeX/bst/l4z_default}
\fi
\ifzdrftbst
  \bibliographystyle{./BiBTeX/bst/l4z_draft}
\fi
\ifzbstepj
  \bibliographystyle{./BiBTeX/bst/l4z_epj}
\fi
\ifzbstnp
  \bibliographystyle{./BiBTeX/bst/l4z_np}
\fi
\ifzbstpl
  \bibliographystyle{./BiBTeX/bst/l4z_pl}
\fi
{\raggedright
\bibliography{./BiBTeX/user/syn,%
              ./BiBTeX/bib/l4z_articles.bib,%
              ./BiBTeX/bib/l4z_books.bib,%
              ./BiBTeX/bib/l4z_conferences.bib,%
              ./BiBTeX/bib/l4z_h1.bib,%
              ./BiBTeX/bib/l4z_misc.bib,%
              ./BiBTeX/bib/l4z_old.bib,%
              ./BiBTeX/bib/l4z_preprints.bib,%
              ./BiBTeX/bib/l4z_replaced.bib,%
              ./BiBTeX/bib/l4z_temporary.bib,%
              ./BiBTeX/bib/l4z_zeus.bib}}
}
\vfill\eject


\begin{table}
\begin{center}
\begin{scriptsize}
\begin{sideways}
\begin{minipage}[t]
{\textheight} \vspace{-1.5cm} 
\begin{center}
\begin{tabular}{|c|c|c|c|c|c|c|c|}
\hline 
$\twoecrbreit$~(GeV) & 1.5 -- 4 & 4 -- 8 & 8 -- 12 & 12 -- 20 & 20 -- 30 & 30 --
45 & 45 -- 100 \\
\hline
\hline

$n_{\rm ch}$=0 & $  12.33 \pm  0.25 ^{+  0.21} _{-  0.23}  $ & 
        $   4.34 \pm  0.09 ^{+  0.65} _{-  0.65}  $ &
	$   1.83 \pm  0.10 ^{+  0.57} _{-  0.56}  $ &
	$   0.81 \pm  0.03 ^{+  0.26} _{-  0.26}  $ &
	$   0.33 \pm  0.03 ^{+  0.10} _{-  0.10}  $ &
	$   0.23 \pm  0.05 ^{+  0.09} _{-  0.09}  $ &
	$   0.17 \pm  0.09 ^{+  0.09} _{-  0.09}  $ \\ 
 \hline 
  1 & $  35.79 \pm  0.46 ^{+  0.18} _{-  0.35}  $ & 
        $  17.44 \pm  0.19 ^{+  0.71} _{-  0.72}  $ &
	$   8.29 \pm  0.20 ^{+  0.78} _{-  0.77}  $ &
	$   4.35 \pm  0.06 ^{+  0.44} _{-  0.44}  $ &
	$   2.23 \pm  0.08 ^{+  0.33} _{-  0.33}  $ &
	$   1.45 \pm  0.13 ^{+  0.45} _{-  0.44}  $ &
	$   0.86 \pm  0.16 ^{+  0.19} _{-  0.20}  $ \\ 
 \hline 
  2 & $  33.31 \pm  0.43 ^{+  0.65} _{-  0.57}  $ & 
        $  27.99 \pm  0.24 ^{+  0.21} _{-  0.21}  $ &
	$  17.14 \pm  0.30 ^{+  0.44} _{-  0.51}  $ &
	$  10.07 \pm  0.09 ^{+  0.38} _{-  0.38}  $ &
	$   5.50 \pm  0.13 ^{+  0.39} _{-  0.38}  $ &
	$   3.38 \pm  0.18 ^{+  0.28} _{-  0.27}  $ &
	$   2.11 \pm  0.25 ^{+  0.38} _{-  0.35}  $ \\ 
 \hline 
  3 & $  14.50 \pm  0.26 ^{+  0.39} _{-  0.39}  $ & 
        $  26.40 \pm  0.23 ^{+  0.50} _{-  0.49}  $ &
	$  24.13 \pm  0.36 ^{+  0.25} _{-  0.37}  $ &
	$  17.08 \pm  0.12 ^{+  0.19} _{-  0.20}  $ &
	$  10.95 \pm  0.19 ^{+  0.29} _{-  0.32}  $ &
	$   7.14 \pm  0.27 ^{+  0.48} _{-  0.42}  $ &
	$   4.45 \pm  0.38 ^{+  0.65} _{-  0.63}  $ \\ 
 \hline 
  4 & $   3.46 \pm  0.12 ^{+  0.18} _{-  0.18}  $ & 
        $  15.27 \pm  0.18 ^{+  0.42} _{-  0.42}  $ &
	$  21.62 \pm  0.34 ^{+  0.65} _{-  0.66}  $ &
	$  19.82 \pm  0.12 ^{+  0.18} _{-  0.27}  $ &
	$  14.65 \pm  0.22 ^{+  0.11} _{-  0.21}  $ &
	$  11.05 \pm  0.34 ^{+  0.30} _{-  0.25}  $ &
	$   7.36 \pm  0.48 ^{+  0.63} _{-  0.60}  $ \\ 
 \hline 
  5 & $   0.54 \pm  0.04 ^{+  0.04} _{-  0.03}  $ & 
        $   6.14 \pm  0.11 ^{+  0.42} _{-  0.40}  $ &
	$  14.72 \pm  0.29 ^{+  0.73} _{-  0.71}  $ &
	$  18.41 \pm  0.12 ^{+  0.43} _{-  0.45}  $ &
	$  16.90 \pm  0.23 ^{+  0.16} _{-  0.15}  $ &
	$  14.03 \pm  0.38 ^{+  0.26} _{-  0.25}  $ &
	$  10.19 \pm  0.56 ^{+  0.42} _{-  0.33}  $ \\ 
 \hline 
  6 & $   0.07 \pm  0.02 ^{+  0.01} _{-  0.01}  $ & 
        $   1.88 \pm  0.06 ^{+  0.14} _{-  0.14}  $ &
	$   7.55 \pm  0.20 ^{+  0.45} _{-  0.40}  $ &
	$  13.36 \pm  0.10 ^{+  0.47} _{-  0.48}  $ &
	$  15.42 \pm  0.22 ^{+  0.30} _{-  0.30}  $ &
	$  14.17 \pm  0.38 ^{+  0.29} _{-  0.36}  $ &
	$  11.66 \pm  0.60 ^{+  0.37} _{-  0.27}  $ \\ 
 \hline 
  7 & $   $ & 
        $   0.43 \pm  0.02 ^{+  0.03} _{-  0.03}  $ &
	$   3.16 \pm  0.13 ^{+  0.22} _{-  0.19}  $ &
	$   8.30 \pm  0.08 ^{+  0.30} _{-  0.26}  $ &
	$  12.86 \pm  0.20 ^{+  0.43} _{-  0.41}  $ &
	$  13.22 \pm  0.36 ^{+  0.60} _{-  0.61}  $ &
	$  11.96 \pm  0.61 ^{+  0.28} _{-  0.16}  $ \\ 
 \hline 
  8 & $   $ & 
        $   0.09 \pm  0.01 ^{+  0.01} _{-  0.01}  $ &
	$   1.13 \pm  0.08 ^{+  0.06} _{-  0.08}  $ &
	$   4.36 \pm  0.06 ^{+  0.17} _{-  0.10}  $ &
	$   8.78 \pm  0.17 ^{+  0.30} _{-  0.30}  $ &
	$  10.96 \pm  0.33 ^{+  0.38} _{-  0.44}  $ &
	$  11.26 \pm  0.59 ^{+  0.66} _{-  0.52}  $ \\ 
 \hline 
  9 & $   $ & 
        $   $ &
	$   0.34 \pm  0.04 ^{+  0.02} _{-  0.02}  $ &
	$   2.09 \pm  0.04 ^{+  0.10} _{-  0.06}  $ &
	$   5.68 \pm  0.13 ^{+  0.28} _{-  0.28}  $ &
	$   8.30 \pm  0.28 ^{+  0.42} _{-  0.41}  $ &
	$  10.10 \pm  0.56 ^{+  0.47} _{-  0.44}  $ \\ 
 \hline 
 10 & $   $ & 
        $   $ &
	$   0.06 \pm  0.01 ^{+  0.01} _{-  0.01}  $ &
	$   0.85 \pm  0.03 ^{+  0.06} _{-  0.04}  $ &
	$   3.27 \pm  0.10 ^{+  0.07} _{-  0.02}  $ &
	$   6.01 \pm  0.25 ^{+  0.22} _{-  0.08}  $ &
	$   7.66 \pm  0.48 ^{+  0.57} _{-  0.65}  $ \\ 
 \hline 
 11 & $   $ & 
        $   $ &
	$   0.02 \pm  0.01 ^{+  0.00} _{-  0.00}  $ &
	$   0.33 \pm  0.02 ^{+  0.04} _{-  0.03}  $ &
	$   1.80 \pm  0.07 ^{+  0.02} _{-  0.05}  $ &
	$   4.00 \pm  0.20 ^{+  0.13} _{-  0.07}  $ &
	$   6.32 \pm  0.44 ^{+  0.18} _{-  0.27}  $ \\ 
 \hline 
 12 & $   $ & 
        $   $ &
	$   $ &
	$   0.12 \pm  0.01 ^{+  0.02} _{-  0.02}  $ &
	$   0.85 \pm  0.05 ^{+  0.08} _{-  0.09}  $ &
	$   2.53 \pm  0.16 ^{+  0.10} _{-  0.09}  $ &
	$   4.80 \pm  0.40 ^{+  0.37} _{-  0.57}  $ \\ 
 \hline 
 13 & $   $ & 
        $   $ &
	$   $ &
	$   $ &
	$   0.45 \pm  0.04 ^{+  0.08} _{-  0.07}  $ &
	$   1.58 \pm  0.12 ^{+  0.05} _{-  0.07}  $ &
	$   3.46 \pm  0.31 ^{+  0.35} _{-  0.41}  $ \\ 
 \hline 
 14 & $   $ & 
        $   $ &
	$   $ &
	$   $ &
	$   0.19 \pm  0.02 ^{+  0.02} _{-  0.02}  $ &
	$   0.98 \pm  0.10 ^{+  0.11} _{-  0.11}  $ &
	$   2.46 \pm  0.26 ^{+  0.07} _{-  0.13}  $ \\ 
 \hline 
 15 & $   $ & 
        $   $ &
	$   $ &
	$   $ &
	$   0.09 \pm  0.02 ^{+  0.02} _{-  0.02}  $ &
	$   0.48 \pm  0.06 ^{+  0.05} _{-  0.02}  $ &
	$   1.77 \pm  0.23 ^{+  0.23} _{-  0.37}  $ \\ 
 \hline 
 16 & $   $ & 
        $   $ &
	$   $ &
	$   $ &
	$   0.03 \pm  0.01 ^{+  0.01} _{-  0.01}  $ &
	$   0.26 \pm  0.05 ^{+  0.03} _{-  0.03}  $ &
	$   1.17 \pm  0.19 ^{+  0.09} _{-  0.11}  $ \\ 
 \hline 
 17 & $   $ & 
        $   $ &
	$   $ &
	$   $ &
	$   0.02 \pm  0.01 ^{+  0.01} _{-  0.02}  $ &
	$   0.14 \pm  0.04 ^{+  0.03} _{-  0.04}  $ &
	$   0.74 \pm  0.15 ^{+  0.13} _{-  0.14}  $ \\ 
 \hline 
 18 & $   $ & 
        $   $ &
	$   $ &
	$   $ &
	$   0.01 \pm  0.01 ^{+  0.00} _{-  0.00}  $ &
	$   0.06 \pm  0.02 ^{+  0.02} _{-  0.02}  $ &
	$   0.65 \pm  0.15 ^{+  0.18} _{-  0.07}  $ \\ 
 \hline 
 19 & $   $ & 
        $   $ &
	$   $ &
	$   $ &
	$   $ &
	$   $ &
	$   0.30 \pm  0.08 ^{+  0.07} _{-  0.11}  $ \\ 
 \hline 
 20 & $   $ & 
        $   $ &
	$   $ &
	$   $ &
	$   $ &
	$   $ &
	$   0.20 \pm  0.06 ^{+  0.17} _{-  0.05}  $ \\ 
 \hline 

\end{tabular}
\caption{Multiplicity distributions $100 \cdot 1/N dN / dn_{ch}$ measured in the current region of the Breit
frame in bins of $\twoecrbreit$. 
 The first errors are statistical and the second are the systematic 
uncertainties.
}
\label{dist_2ebr}
\end{center}
\end{minipage}
\end{sideways}
\end{scriptsize}
\end{center}
\end{table}

\begin{table}
\begin{scriptsize}
\begin{center}
\begin{tabular}{|c|c|c|c|c|}
\hline 
$\meff$~(GeV) & 1.5 -- 4 & 4 -- 8 & 8 -- 12 & 12 -- 20 \\
\hline
\hline
 
 $n_{\rm ch}$=0 & $   1.90 \pm  0.05 ^{+  0.34} _{-  0.34}  $ & 
        $   0.23 \pm  0.04 ^{+  0.07} _{-  0.07}  $ &
	$   0.04 \pm  0.02 ^{+  0.02} _{-  0.01}  $ &
	$   0.02 \pm  0.02 ^{+  0.01} _{-  0.01}  $ \\ 
 \hline 
  1 & $   9.81 \pm  0.13 ^{+  0.47} _{-  0.50}  $ & 
        $   1.68 \pm  0.12 ^{+  0.24} _{-  0.22}  $ &
	$   0.40 \pm  0.05 ^{+  0.02} _{-  0.02}  $ &
	$   0.13 \pm  0.05 ^{+  0.02} _{-  0.02}  $ \\ 
 \hline 
  2 & $  21.73 \pm  0.20 ^{+  0.28} _{-  0.32}  $ & 
        $   5.57 \pm  0.24 ^{+  0.31} _{-  0.26}  $ &
	$   1.26 \pm  0.09 ^{+  0.04} _{-  0.07}  $ &
	$   0.56 \pm  0.13 ^{+  0.21} _{-  0.21}  $ \\ 
 \hline 
  3 & $  28.19 \pm  0.23 ^{+  0.28} _{-  0.29}  $ & 
        $  12.12 \pm  0.37 ^{+  0.35} _{-  0.24}  $ &
	$   3.58 \pm  0.16 ^{+  0.16} _{-  0.14}  $ &
	$   1.49 \pm  0.18 ^{+  0.19} _{-  0.17}  $ \\ 
 \hline 
  4 & $  21.12 \pm  0.20 ^{+  0.15} _{-  0.14}  $ & 
        $  17.68 \pm  0.47 ^{+  0.31} _{-  0.40}  $ &
	$   6.73 \pm  0.23 ^{+  0.09} _{-  0.12}  $ &
	$   3.33 \pm  0.30 ^{+  0.26} _{-  0.30}  $ \\ 
 \hline 
  5 & $  10.93 \pm  0.14 ^{+  0.32} _{-  0.25}  $ & 
        $  19.78 \pm  0.51 ^{+  0.21} _{-  0.29}  $ &
	$  11.07 \pm  0.31 ^{+  0.34} _{-  0.37}  $ &
	$   5.64 \pm  0.39 ^{+  0.27} _{-  0.18}  $ \\ 
 \hline 
  6 & $   4.40 \pm  0.09 ^{+  0.13} _{-  0.08}  $ & 
        $  17.11 \pm  0.48 ^{+  0.69} _{-  0.69}  $ &
	$  14.17 \pm  0.37 ^{+  0.11} _{-  0.20}  $ &
	$   8.38 \pm  0.50 ^{+  0.24} _{-  0.12}  $ \\ 
 \hline 
  7 & $   1.43 \pm  0.05 ^{+  0.05} _{-  0.03}  $ & 
        $  12.41 \pm  0.41 ^{+  0.17} _{-  0.18}  $ &
	$  16.01 \pm  0.40 ^{+  0.39} _{-  0.42}  $ &
	$  10.47 \pm  0.58 ^{+  0.40} _{-  0.35}  $ \\ 
 \hline 
  8 & $   0.38 \pm  0.02 ^{+  0.02} _{-  0.01}  $ & 
        $   7.27 \pm  0.31 ^{+  0.25} _{-  0.24}  $ &
	$  14.40 \pm  0.38 ^{+  0.46} _{-  0.46}  $ &
	$  12.16 \pm  0.66 ^{+  0.43} _{-  0.42}  $ \\ 
 \hline 
  9 & $   0.09 \pm  0.01 ^{+  0.01} _{-  0.01}  $ & 
        $   3.51 \pm  0.23 ^{+  0.47} _{-  0.46}  $ &
	$  11.74 \pm  0.35 ^{+  0.29} _{-  0.26}  $ &
	$  12.75 \pm  0.68 ^{+  0.27} _{-  0.17}  $ \\ 
 \hline 
 10 & $   $ & 
        $   1.63 \pm  0.15 ^{+  0.17} _{-  0.16}  $ &
	$   8.46 \pm  0.30 ^{+  0.33} _{-  0.29}  $ &
	$  11.80 \pm  0.69 ^{+  0.28} _{-  0.26}  $ \\ 
 \hline 
 11 & $   $ & 
        $   0.66 \pm  0.09 ^{+  0.09} _{-  0.08}  $ &
	$   5.50 \pm  0.24 ^{+  0.25} _{-  0.21}  $ &
	$  10.03 \pm  0.63 ^{+  0.33} _{-  0.12}  $ \\ 
 \hline 
 12 & $   $ & 
        $   0.26 \pm  0.06 ^{+  0.05} _{-  0.05}  $ &
	$   3.16 \pm  0.18 ^{+  0.32} _{-  0.31}  $ &
	$   7.74 \pm  0.57 ^{+  0.25} _{-  0.34}  $ \\ 
 \hline 
 13 & $   $ & 
        $   0.04 \pm  0.02 ^{+  0.01} _{-  0.01}  $ &
	$   1.76 \pm  0.13 ^{+  0.24} _{-  0.20}  $ &
	$   5.56 \pm  0.46 ^{+  0.21} _{-  0.48}  $ \\ 
 \hline 
 14 & $   $ & 
        $   0.03 \pm  0.02 ^{+  0.01} _{-  0.01}  $ &
	$   0.94 \pm  0.09 ^{+  0.13} _{-  0.11}  $ &
	$   4.02 \pm  0.40 ^{+  0.33} _{-  0.21}  $ \\ 
 \hline 
 15 & $   $ & 
        $   0.01 \pm  0.01 ^{+  0.01} _{-  0.01}  $ &
	$   0.41 \pm  0.06 ^{+  0.13} _{-  0.12}  $ &
	$   2.50 \pm  0.31 ^{+  0.11} _{-  0.22}  $ \\ 
 \hline 
 16 & $   $ & 
        $   $ &
	$   0.19 \pm  0.04 ^{+  0.04} _{-  0.04}  $ &
	$   1.44 \pm  0.25 ^{+  0.07} _{-  0.14}  $ \\ 
 \hline 
 17 & $   $ & 
        $   $ &
	$   0.10 \pm  0.03 ^{+  0.03} _{-  0.05}  $ &
	$   0.89 \pm  0.19 ^{+  0.21} _{-  0.23}  $ \\ 
 \hline 
 18 & $   $ & 
        $   $ &
	$   0.05 \pm  0.02 ^{+  0.03} _{-  0.02}  $ &
	$   0.52 \pm  0.14 ^{+  0.13} _{-  0.13}  $ \\ 
 \hline 
 19 & $   $ & 
        $   $ &
	$   0.01 \pm  0.01 ^{+  0.00} _{-  0.00}  $ &
	$   0.22 \pm  0.09 ^{+  0.11} _{-  0.11}  $ \\ 
 \hline 
 20 & $   $ & 
        $   $ &
	$   $ &
	$   0.19 \pm  0.09 ^{+  0.13} _{-  0.10}  $ \\ 
 \hline 

\end{tabular}
\caption{Multiplicity distributions $100 \cdot 1/N dN / dn_{ch} $ measured in the current region of the Breit
frame in bins of $\meff$. 
 The first errors are statistical and the second are the systematic
uncertainties.
}
\label{dist_brmeff}
\end{center}
\end{scriptsize}
\end{table}

\begin{table}
\begin{scriptsize}
\begin{center}
\begin{tabular}{|c|c|c|c|}
\hline 
$W$~(GeV) & 70 -- 100 & 100 -- 150 & 150 -- 225  \\
\hline
\hline
 
 $n_{\rm ch}$=0 & $   0.09 \pm  0.03 ^{+  0.01} _{-  0.01}  $ & 
        $   0.11 \pm  0.03 ^{+  0.02} _{-  0.02}  $ &
	$   0.09 \pm  0.02 ^{+  0.05} _{-  0.04}  $ \\ 
 \hline 
  1 & $   0.37 \pm  0.04 ^{+  0.13} _{-  0.13}  $ & 
        $   0.26 \pm  0.03 ^{+  0.10} _{-  0.10}  $ &
	$   0.25 \pm  0.04 ^{+  0.17} _{-  0.17}  $ \\ 
 \hline 
  2 & $   1.50 \pm  0.07 ^{+  0.28} _{-  0.28}  $ & 
        $   1.02 \pm  0.06 ^{+  0.23} _{-  0.23}  $ &
	$   0.81 \pm  0.05 ^{+  0.19} _{-  0.19}  $ \\ 
 \hline 
  3 & $   2.58 \pm  0.09 ^{+  0.36} _{-  0.36}  $ & 
        $   1.55 \pm  0.06 ^{+  0.17} _{-  0.17}  $ &
	$   0.99 \pm  0.05 ^{+  0.18} _{-  0.17}  $ \\ 
 \hline 
  4 & $   5.22 \pm  0.13 ^{+  0.52} _{-  0.52}  $ & 
        $   3.60 \pm  0.09 ^{+  0.69} _{-  0.69}  $ &
	$   2.43 \pm  0.08 ^{+  0.66} _{-  0.66}  $ \\ 
 \hline 
  5 & $   6.47 \pm  0.14 ^{+  0.39} _{-  0.38}  $ & 
        $   4.43 \pm  0.10 ^{+  0.14} _{-  0.13}  $ &
	$   3.00 \pm  0.08 ^{+  0.07} _{-  0.06}  $ \\ 
 \hline 
  6 & $   8.97 \pm  0.16 ^{+  0.58} _{-  0.58}  $ & 
        $   6.71 \pm  0.12 ^{+  0.85} _{-  0.85}  $ &
	$   4.72 \pm  0.10 ^{+  0.80} _{-  0.80}  $ \\ 
 \hline 
  7 & $   9.57 \pm  0.17 ^{+  0.12} _{-  0.09}  $ & 
        $   7.40 \pm  0.13 ^{+  0.19} _{-  0.18}  $ &
	$   5.51 \pm  0.11 ^{+  0.24} _{-  0.23}  $ \\ 
 \hline 
  8 & $   9.89 \pm  0.16 ^{+  0.78} _{-  0.78}  $ & 
        $   8.41 \pm  0.13 ^{+  0.89} _{-  0.89}  $ &
	$   6.82 \pm  0.12 ^{+  0.87} _{-  0.87}  $ \\ 
 \hline 
  9 & $   9.53 \pm  0.16 ^{+  0.53} _{-  0.53}  $ & 
        $   8.45 \pm  0.13 ^{+  0.51} _{-  0.51}  $ &
	$   7.18 \pm  0.12 ^{+  0.57} _{-  0.57}  $ \\ 
 \hline 
 10 & $   8.73 \pm  0.16 ^{+  0.70} _{-  0.70}  $ & 
        $   8.40 \pm  0.13 ^{+  0.85} _{-  0.85}  $ &
	$   7.61 \pm  0.13 ^{+  0.74} _{-  0.74}  $ \\ 
 \hline 
 11 & $   7.82 \pm  0.15 ^{+  0.41} _{-  0.41}  $ & 
        $   8.05 \pm  0.13 ^{+  0.40} _{-  0.41}  $ &
	$   7.54 \pm  0.13 ^{+  0.57} _{-  0.57}  $ \\ 
 \hline 
 12 & $   6.63 \pm  0.14 ^{+  0.16} _{-  0.16}  $ & 
        $   7.31 \pm  0.12 ^{+  0.49} _{-  0.49}  $ &
	$   7.41 \pm  0.13 ^{+  0.58} _{-  0.58}  $ \\ 
 \hline 
 13 & $   5.50 \pm  0.13 ^{+  0.11} _{-  0.12}  $ & 
        $   6.46 \pm  0.12 ^{+  0.15} _{-  0.15}  $ &
	$   6.82 \pm  0.12 ^{+  0.30} _{-  0.30}  $ \\ 
 \hline 
 14 & $   4.32 \pm  0.12 ^{+  0.12} _{-  0.15}  $ & 
        $   5.53 \pm  0.11 ^{+  0.03} _{-  0.02}  $ &
	$   6.33 \pm  0.12 ^{+  0.31} _{-  0.31}  $ \\ 
 \hline 
 15 & $   3.45 \pm  0.11 ^{+  0.25} _{-  0.25}  $ & 
        $   4.72 \pm  0.10 ^{+  0.19} _{-  0.19}  $ &
	$   5.69 \pm  0.11 ^{+  0.05} _{-  0.05}  $ \\ 
 \hline 
 16 & $   2.58 \pm  0.09 ^{+  0.37} _{-  0.37}  $ & 
        $   3.90 \pm  0.10 ^{+  0.27} _{-  0.27}  $ &
	$   4.93 \pm  0.11 ^{+  0.00} _{-  0.03}  $ \\ 
 \hline 
 17 & $   1.96 \pm  0.09 ^{+  0.44} _{-  0.44}  $ & 
        $   3.16 \pm  0.09 ^{+  0.44} _{-  0.44}  $ &
	$   4.22 \pm  0.10 ^{+  0.30} _{-  0.30}  $ \\ 
 \hline 
 18 & $   1.46 \pm  0.08 ^{+  0.45} _{-  0.45}  $ & 
        $   2.58 \pm  0.08 ^{+  0.45} _{-  0.45}  $ &
	$   3.48 \pm  0.09 ^{+  0.30} _{-  0.30}  $ \\ 
 \hline 
 19 & $   1.05 \pm  0.07 ^{+  0.35} _{-  0.35}  $ & 
        $   1.98 \pm  0.08 ^{+  0.51} _{-  0.51}  $ &
	$   2.95 \pm  0.09 ^{+  0.37} _{-  0.37}  $ \\ 
 \hline 
 20 & $   0.74 \pm  0.06 ^{+  0.34} _{-  0.34}  $ & 
        $   1.50 \pm  0.07 ^{+  0.51} _{-  0.51}  $ &
	$   2.32 \pm  0.08 ^{+  0.45} _{-  0.45}  $ \\ 
 \hline 
 21 & $   0.51 \pm  0.06 ^{+  0.25} _{-  0.25}  $ & 
        $   1.14 \pm  0.06 ^{+  0.48} _{-  0.48}  $ &
	$   1.91 \pm  0.08 ^{+  0.52} _{-  0.52}  $ \\ 
 \hline 
 22 & $   0.34 \pm  0.05 ^{+  0.19} _{-  0.19}  $ & 
        $   0.90 \pm  0.06 ^{+  0.38} _{-  0.38}  $ &
	$   1.51 \pm  0.07 ^{+  0.48} _{-  0.48}  $ \\ 
 \hline 
 23 & $   0.25 \pm  0.05 ^{+  0.16} _{-  0.16}  $ & 
        $   0.65 \pm  0.05 ^{+  0.36} _{-  0.36}  $ &
	$   1.22 \pm  0.07 ^{+  0.47} _{-  0.47}  $ \\ 
 \hline 
 24 & $   0.17 \pm  0.04 ^{+  0.12} _{-  0.12}  $ & 
        $   0.46 \pm  0.05 ^{+  0.27} _{-  0.27}  $ &
	$   0.94 \pm  0.06 ^{+  0.41} _{-  0.41}  $ \\ 
 \hline 
 25 & $   0.10 \pm  0.04 ^{+  0.08} _{-  0.08}  $ & 
        $   0.36 \pm  0.05 ^{+  0.23} _{-  0.23}  $ &
	$   0.74 \pm  0.06 ^{+  0.40} _{-  0.40}  $ \\ 
 \hline 
 26 & $   0.08 \pm  0.03 ^{+  0.06} _{-  0.06}  $ & 
        $   0.28 \pm  0.04 ^{+  0.20} _{-  0.20}  $ &
	$   0.58 \pm  0.05 ^{+  0.31} _{-  0.31}  $ \\ 
 \hline 
 27 & $   0.05 \pm  0.03 ^{+  0.04} _{-  0.04}  $ & 
        $   0.19 \pm  0.04 ^{+  0.15} _{-  0.15}  $ &
	$   0.45 \pm  0.05 ^{+  0.27} _{-  0.27}  $ \\ 
 \hline 
 28 & $   0.03 \pm  0.02 ^{+  0.03} _{-  0.03}  $ & 
        $   0.15 \pm  0.04 ^{+  0.12} _{-  0.12}  $ &
	$   0.36 \pm  0.05 ^{+  0.23} _{-  0.23}  $ \\ 
 \hline 
 29 & $   0.02 \pm  0.02 ^{+  0.01} _{-  0.02}  $ & 
        $   0.10 \pm  0.04 ^{+  0.08} _{-  0.08}  $ &
	$   0.29 \pm  0.05 ^{+  0.20} _{-  0.20}  $ \\ 
 \hline 
 30 & $   0.01 \pm  0.03 ^{+  0.01} _{-  0.02}  $ & 
        $   0.07 \pm  0.02 ^{+  0.05} _{-  0.05}  $ &
	$   0.22 \pm  0.05 ^{+  0.17} _{-  0.17}  $ \\ 
 \hline 
 31 & $   $ & 
        $   0.05 \pm  0.03 ^{+  0.04} _{-  0.04}  $ &
	$   0.17 \pm  0.04 ^{+  0.12} _{-  0.12}  $ \\ 
 \hline 
 32 & $   $ & 
        $   0.04 \pm  0.03 ^{+  0.03} _{-  0.03}  $ &
	$   0.13 \pm  0.05 ^{+  0.11} _{-  0.11}  $ \\ 
 \hline 
 33 & $   $ & 
        $   0.03 \pm  0.03 ^{+  0.03} _{-  0.03}  $ &
	$   0.10 \pm  0.04 ^{+  0.08} _{-  0.08}  $ \\ 
 \hline 
 34 & $   $ & 
        $   0.01 \pm  0.02 ^{+  0.01} _{-  0.01}  $ &
	$   0.08 \pm  0.03 ^{+  0.06} _{-  0.06}  $ \\ 
 \hline 
 35 & $   $ & 
        $   0.01 \pm  0.01 ^{+  0.01} _{-  0.01}  $ &
	$   0.05 \pm  0.03 ^{+  0.05} _{-  0.07}  $ \\ 
 \hline 
 36 & $   $ & 
        $   $ &
	$   0.05 \pm  0.03 ^{+  0.05} _{-  0.04}  $ \\ 
 \hline 
 37 & $   $ & 
        $   $ &
	$   0.03 \pm  0.03 ^{+  0.03} _{-  0.03}  $ \\ 
 \hline 
 38 & $   $ & 
        $   $ &
	$   0.02 \pm  0.02 ^{+  0.02} _{-  0.02}  $ \\  
 \hline 

\end{tabular}
\caption{Multiplicity distributions $100 \cdot 1/N dN / dn_{ch} $ measured in the current region of the HCM
 in bins of $W$.
 The first errors are statistical and the second are the systematic
uncertainties.
}
\label{dist_whcm}
\end{center}
\end{scriptsize}
\end{table}

\begin{table}
\begin{scriptsize}
\begin{center}
\begin{tabular}{|c|c|c|c|c|c|}
\hline 
$\meff$~(GeV) & 1.5 -- 4 & 4 -- 8 & 8 -- 12 & 12 -- 20 & 20 -- 30 \\
\hline
\hline
 
 $n_{\rm ch}$=0 & $   1.44 \pm  0.09 ^{+  0.58} _{-  0.58}  $ & 
        $   0.12 \pm  0.01 ^{+  0.04} _{-  0.04}  $ &
	$   0.02 \pm  0.01 ^{+  0.00} _{-  0.00}  $ &
	$   $ &
	$   $ \\ 
 \hline 
  1 & $   6.59 \pm  0.22 ^{+  0.74} _{-  0.74}  $ & 
        $   1.06 \pm  0.04 ^{+  0.05} _{-  0.06}  $ &
	$   0.18 \pm  0.02 ^{+  0.03} _{-  0.03}  $ &
	$   0.05 \pm  0.01 ^{+  0.01} _{-  0.01}  $ &
	$   $ \\ 
 \hline 
  2 & $  16.69 \pm  0.40 ^{+  1.54} _{-  1.50}  $ & 
        $   3.83 \pm  0.09 ^{+  0.13} _{-  0.13}  $ &
	$   0.76 \pm  0.04 ^{+  0.01} _{-  0.04}  $ &
	$   0.17 \pm  0.02 ^{+  0.01} _{-  0.01}  $ &
	$   0.03 \pm  0.02 ^{+  0.01} _{-  0.01}  $ \\ 
 \hline 
  3 & $  23.29 \pm  0.49 ^{+  0.27} _{-  0.21}  $ & 
        $   8.91 \pm  0.14 ^{+  0.27} _{-  0.22}  $ &
	$   2.27 \pm  0.08 ^{+  0.03} _{-  0.10}  $ &
	$   0.54 \pm  0.04 ^{+  0.01} _{-  0.03}  $ &
	$   0.13 \pm  0.04 ^{+  0.05} _{-  0.03}  $ \\ 
 \hline 
  4 & $  24.20 \pm  0.52 ^{+  1.26} _{-  1.28}  $ & 
        $  15.01 \pm  0.19 ^{+  0.81} _{-  0.82}  $ &
	$   5.15 \pm  0.12 ^{+  0.32} _{-  0.38}  $ &
	$   1.32 \pm  0.06 ^{+  0.08} _{-  0.09}  $ &
	$   0.31 \pm  0.06 ^{+  0.08} _{-  0.06}  $ \\ 
 \hline 
  5 & $  15.49 \pm  0.41 ^{+  0.87} _{-  0.86}  $ & 
        $  18.40 \pm  0.21 ^{+  0.14} _{-  0.21}  $ &
	$   8.84 \pm  0.16 ^{+  0.46} _{-  0.48}  $ &
	$   2.81 \pm  0.09 ^{+  0.19} _{-  0.20}  $ &
	$   0.78 \pm  0.10 ^{+  0.10} _{-  0.07}  $ \\ 
 \hline 
  6 & $   8.07 \pm  0.29 ^{+  0.34} _{-  0.32}  $ & 
        $  18.13 \pm  0.22 ^{+  0.71} _{-  0.71}  $ &
	$  12.51 \pm  0.19 ^{+  0.64} _{-  0.65}  $ &
	$   4.81 \pm  0.12 ^{+  0.41} _{-  0.43}  $ &
	$   1.49 \pm  0.13 ^{+  0.13} _{-  0.07}  $ \\ 
 \hline 
  7 & $   3.03 \pm  0.17 ^{+  0.16} _{-  0.18}  $ & 
        $  14.32 \pm  0.19 ^{+  0.32} _{-  0.34}  $ &
	$  14.61 \pm  0.21 ^{+  0.79} _{-  0.79}  $ &
	$   7.30 \pm  0.15 ^{+  0.68} _{-  0.69}  $ &
	$   2.43 \pm  0.18 ^{+  0.18} _{-  0.23}  $ \\ 
 \hline 
  8 & $   0.95 \pm  0.09 ^{+  0.13} _{-  0.12}  $ & 
        $   9.62 \pm  0.16 ^{+  0.26} _{-  0.24}  $ &
	$  14.62 \pm  0.21 ^{+  0.60} _{-  0.59}  $ &
	$   9.38 \pm  0.18 ^{+  0.68} _{-  0.69}  $ &
	$   3.91 \pm  0.22 ^{+  0.65} _{-  0.59}  $ \\ 
 \hline 
  9 & $   0.20 \pm  0.04 ^{+  0.06} _{-  0.07}  $ & 
        $   5.54 \pm  0.12 ^{+  0.50} _{-  0.49}  $ &
	$  12.97 \pm  0.20 ^{+  0.20} _{-  0.16}  $ &
	$  11.03 \pm  0.19 ^{+  0.85} _{-  0.86}  $ &
	$   5.56 \pm  0.28 ^{+  0.95} _{-  0.93}  $ \\ 
 \hline 
 10 & $   0.05 \pm  0.02 ^{+  0.03} _{-  0.01}  $ & 
        $   2.82 \pm  0.09 ^{+  0.34} _{-  0.33}  $ &
	$  10.32 \pm  0.19 ^{+  0.28} _{-  0.27}  $ &
	$  11.53 \pm  0.20 ^{+  0.63} _{-  0.63}  $ &
	$   6.97 \pm  0.31 ^{+  1.10} _{-  1.13}  $ \\ 
 \hline 
 11 & $   $ & 
        $   1.36 \pm  0.06 ^{+  0.28} _{-  0.28}  $ &
	$   7.24 \pm  0.16 ^{+  0.46} _{-  0.44}  $ &
	$  11.18 \pm  0.20 ^{+  0.59} _{-  0.59}  $ &
	$   8.01 \pm  0.33 ^{+  1.23} _{-  1.25}  $ \\ 
 \hline 
 12 & $   $ & 
        $   0.57 \pm  0.04 ^{+  0.13} _{-  0.13}  $ &
	$   4.74 \pm  0.13 ^{+  0.71} _{-  0.70}  $ &
	$  10.07 \pm  0.20 ^{+  0.20} _{-  0.21}  $ &
	$   8.48 \pm  0.35 ^{+  1.05} _{-  1.04}  $ \\ 
 \hline 
 13 & $   $ & 
        $   0.19 \pm  0.03 ^{+  0.07} _{-  0.07}  $ &
	$   2.73 \pm  0.10 ^{+  0.59} _{-  0.58}  $ &
	$   8.44 \pm  0.18 ^{+  0.27} _{-  0.24}  $ &
	$   9.26 \pm  0.36 ^{+  1.36} _{-  1.36}  $ \\ 
 \hline 
 14 & $   $ & 
        $   0.08 \pm  0.02 ^{+  0.02} _{-  0.02}  $ &
	$   1.51 \pm  0.08 ^{+  0.38} _{-  0.37}  $ &
	$   6.58 \pm  0.16 ^{+  0.52} _{-  0.51}  $ &
	$   8.90 \pm  0.38 ^{+  0.72} _{-  0.71}  $ \\ 
 \hline 
 15 & $   $ & 
        $   0.03 \pm  0.01 ^{+  0.00} _{-  0.01}  $ &
	$   0.83 \pm  0.06 ^{+  0.28} _{-  0.28}  $ &
	$   4.92 \pm  0.15 ^{+  0.60} _{-  0.59}  $ &
	$   8.09 \pm  0.36 ^{+  0.37} _{-  0.35}  $ \\ 
 \hline 
 16 & $   $ & 
        $   $ &
	$   0.35 \pm  0.04 ^{+  0.14} _{-  0.14}  $ &
	$   3.53 \pm  0.13 ^{+  0.66} _{-  0.65}  $ &
	$   7.57 \pm  0.36 ^{+  0.21} _{-  0.21}  $ \\ 
 \hline 
 17 & $   $ & 
        $   $ &
	$   0.20 \pm  0.03 ^{+  0.08} _{-  0.08}  $ &
	$   2.43 \pm  0.11 ^{+  0.64} _{-  0.63}  $ &
	$   6.41 \pm  0.35 ^{+  0.47} _{-  0.45}  $ \\ 
 \hline 
 18 & $   $ & 
        $   $ &
	$   0.08 \pm  0.02 ^{+  0.03} _{-  0.03}  $ &
	$   1.56 \pm  0.10 ^{+  0.56} _{-  0.56}  $ &
	$   5.17 \pm  0.33 ^{+  0.86} _{-  0.90}  $ \\ 
 \hline 
 19 & $   $ & 
        $   $ &
	$   0.03 \pm  0.02 ^{+  0.02} _{-  0.02}  $ &
	$   0.96 \pm  0.08 ^{+  0.36} _{-  0.36}  $ &
	$   4.08 \pm  0.29 ^{+  0.65} _{-  0.71}  $ \\ 
 \hline 
 20 & $   $ & 
        $   $ &
	$   0.01 \pm  0.01 ^{+  0.01} _{-  0.01}  $ &
	$   0.59 \pm  0.06 ^{+  0.25} _{-  0.25}  $ &
	$   3.47 \pm  0.29 ^{+  0.96} _{-  0.95}  $ \\ 
 \hline 
 21 & $   $ & 
        $   $ &
	$   $ &
	$   0.36 \pm  0.05 ^{+  0.18} _{-  0.18}  $ &
	$   2.63 \pm  0.27 ^{+  0.99} _{-  0.98}  $ \\ 
 \hline 
 22 & $   $ & 
        $   $ &
	$   $ &
	$   0.18 \pm  0.04 ^{+  0.12} _{-  0.11}  $ &
	$   1.83 \pm  0.24 ^{+  0.72} _{-  0.72}  $ \\ 
 \hline 
 23 & $   $ & 
        $   $ &
	$   $ &
	$   0.12 \pm  0.04 ^{+  0.09} _{-  0.08}  $ &
	$   1.39 \pm  0.23 ^{+  0.77} _{-  0.75}  $ \\ 
 \hline 
 24 & $   $ & 
        $   $ &
	$   $ &
	$   0.07 \pm  0.02 ^{+  0.04} _{-  0.04}  $ &
	$   0.96 \pm  0.20 ^{+  0.62} _{-  0.63}  $ \\ 
 \hline 
 25 & $   $ & 
        $   $ &
	$   $ &
	$   0.04 \pm  0.02 ^{+  0.02} _{-  0.02}  $ &
	$   0.66 \pm  0.20 ^{+  0.50} _{-  0.52}  $ \\ 
 \hline 
 26 & $   $ & 
        $   $ &
	$   $ &
	$   0.02 \pm  0.02 ^{+  0.01} _{-  0.02}  $ &
	$   0.53 \pm  0.16 ^{+  0.33} _{-  0.33}  $ \\ 
 \hline 
 27 & $   $ & 
        $   $ &
	$   $ &
	$   $ &
	$   0.34 \pm  0.16 ^{+  0.27} _{-  0.22}  $ \\ 
 \hline 
 28 & $   $ & 
        $   $ &
	$   $ &
	$   $ &
	$   0.28 \pm  0.20 ^{+  0.26} _{-  0.26}  $ \\ 
 \hline 
 29 & $   $ & 
        $   $ &
	$   $ &
	$   $ &
	$   0.13 \pm  0.11 ^{+  0.11} _{-  0.13}  $ \\ 
 \hline 
 30 & $   $ & 
        $   $ &
	$   $ &
	$   $ &
	$   0.09 \pm  0.07 ^{+  0.07} _{-  0.12}  $ \\ 
 \hline 

\end{tabular}
\caption{Multiplicity distributions $100 \cdot 1/N dN / dn_{ch}$ measured in the current region of the HCM
frame in bins of $\meff$. 
 The first errors are statistical and the second are the systematic
uncertainties.
}
\label{dist_hcmmeff}
\end{center}
\end{scriptsize}
\end{table}

\begin{table}
\begin{center}
\begin{tabular}{|c|c|c|}
\hline 
$ \twoecrbreit$~(GeV)  & $\nch$ ($K^0, \Lambda$ stable) &
$\nch$ ($K^0, \Lambda$ decay) \\
\hline 
\hline
 
 2.9 & $ 1.50 \pm 0.01 ^{+ 0.04} _{- 0.00} $ & 
        $ 1.63 \pm 0.01 ^{+ 0.05} _{- 0.01} $ \\ 
 5.9 & $ 2.39 \pm 0.01 ^{+ 0.09} _{- 0.05} $ & 
        $ 2.60 \pm 0.01 ^{+ 0.11} _{- 0.05} $ \\ 
 9.6 & $ 3.27 \pm 0.01 ^{+ 0.12} _{- 0.08} $ & 
        $ 3.55 \pm 0.01 ^{+ 0.15} _{- 0.08} $ \\ 
  14.8 & $ 4.17 \pm 0.01 ^{+ 0.10} _{- 0.05} $ & 
        $ 4.53 \pm 0.01 ^{+ 0.12} _{- 0.05} $ \\ 
  23.8 & $ 5.22 \pm 0.01 ^{+ 0.06} _{- 0.05} $ & 
        $ 5.67 \pm 0.01 ^{+ 0.08} _{- 0.05} $ \\ 
  35.6 & $ 6.19 \pm 0.03 ^{+ 0.08} _{- 0.07} $ & 
        $ 6.68 \pm 0.03 ^{+ 0.08} _{- 0.08} $ \\ 
  58.1 & $ 7.46 \pm 0.06 ^{+ 0.14} _{- 0.16} $ & 
        $ 8.04 \pm 0.06 ^{+ 0.16} _{- 0.18} $ \\ 
 \hline 
 \hline 

$W$~(GeV) & $\nch$ ($K^0, \Lambda$ stable) &
$\nch$ ($K^0, \Lambda$ decay) \\
\hline 
\hline
 
  84.6 & $ 8.72 \pm 0.02 ^{+ 0.22} _{- 0.22} $ & 
        $ 9.58 \pm 0.02 ^{+ 0.33} _{- 0.33} $ \\ 
 123.8 & $  10.04 \pm 0.02 ^{+ 0.44} _{- 0.44} $ & 
        $  11.07 \pm 0.02 ^{+ 0.61} _{- 0.61} $ \\ 
 184.5 & $  11.40 \pm 0.03 ^{+ 0.58} _{- 0.58} $ & 
        $  12.59 \pm 0.03 ^{+ 0.79} _{- 0.79} $ \\ 
 \hline 

\end{tabular}
\caption{
Mean charged multiplicity, $\nch$, measured in the current
region of the Breit frame as a function of $\twoecrbreit$
and in the current fragmentation region
of the HCM frame as a function of $W$.
 The first errors are statistical and the second are the systematic
uncertainties.
}
\label{ave_2e}
\end{center}
\end{table}

\begin{table}
\begin{center}
\begin{tabular}{|c|c|c|}

\hline
\multicolumn{3}{|c|}{ Current region of the Breit frame} \\
\hline 
$ \meff$~(GeV)  & $\nch$ ($K^0, \Lambda$ stable) &
$\nch$ ($K^0, \Lambda$ decay) \\
\hline 
\hline
 
 2.4 & $ 2.91 \pm 0.01 ^{+ 0.04} _{- 0.03} $ & 
        $ 3.17 \pm 0.01 ^{+ 0.04} _{- 0.03} $ \\ 
 \hline 
 5.2 & $ 4.82 \pm 0.02 ^{+ 0.03} _{- 0.03} $ & 
        $ 5.26 \pm 0.02 ^{+ 0.03} _{- 0.03} $ \\ 
 \hline 
 9.4 & $ 6.85 \pm 0.02 ^{+ 0.07} _{- 0.05} $ & 
        $ 7.45 \pm 0.02 ^{+ 0.08} _{- 0.06} $ \\ 
 \hline 
  14.4 & $ 8.60 \pm 0.06 ^{+ 0.07} _{- 0.09} $ & 
        $ 9.29 \pm 0.06 ^{+ 0.14} _{- 0.09} $ \\ 
 \hline 
 \hline

\multicolumn{3}{|c|}{ Current region of the HCM frame} \\
\hline
$\meff$~(GeV) & $\nch$ ($K^0, \Lambda$ stable) &
$\nch$ ($K^0, \Lambda$ decay) \\
\hline 
\hline
 
 3.1 & $ 3.38 \pm 0.02 ^{+ 0.03} _{- 0.02} $ & 
        $ 3.64 \pm 0.02 ^{+ 0.05} _{- 0.02} $ \\ 
 \hline 
 5.9 & $ 5.33 \pm 0.01 ^{+ 0.07} _{- 0.07} $ & 
        $ 5.77 \pm 0.01 ^{+ 0.09} _{- 0.09} $ \\ 
 \hline 
 9.8 & $ 7.37 \pm 0.02 ^{+ 0.13} _{- 0.12} $ & 
        $ 8.05 \pm 0.01 ^{+ 0.19} _{- 0.19} $ \\ 
 \hline 
  15.1 & $ 9.86 \pm 0.02 ^{+ 0.23} _{- 0.23} $ & 
        $  10.84 \pm 0.02 ^{+ 0.36} _{- 0.36} $ \\ 
 \hline 
  23.5 & $  12.83 \pm 0.06 ^{+ 0.53} _{- 0.54} $ & 
        $  14.17 \pm 0.05 ^{+ 0.80} _{- 0.80} $ \\ 
 \hline 

\end{tabular}
\caption{
Mean charged multiplicity, $\nch$, measured in the current
region of the Breit frame and in the current fragmentation region
of the HCM frame as a function of $\meff$.
 The first errors are statistical and the second are the systematic
uncertainties.
}\label{ave_meff}
\end{center}
\end{table}

%
%
\end{document}